\documentclass[aps,prb,preprint]{revtex4}%
\usepackage{amsfonts}
\usepackage{amsmath}
\usepackage{amssymb}
\usepackage{verbatim}
\usepackage{graphicx}%
\providecommand{\U}[1]{\protect\rule{.1in}{.1in}}

\begin{document}
\title{Phonon driven transport in amorphous semiconductors: Transition probabilities}
\author{Ming-Liang Zhang}
\affiliation{Department of Physics and Astronomy, Ohio University, Athens, Ohio 45701}
\author{D. A. Drabold}
\affiliation{Trinity College, Cambridge, CB2 1TQ, United Kingdom and Department of Physics
and Astronomy, Ohio University, Athens, Ohio 45701}

\begin{abstract}
Inspired by Holstein's work on small polaron hopping, the evolution equations
of localized states and extended states in presence of atomic vibrations are
derived for an amorphous semiconductor. The transition probabilities are
obtained for four types of transitions: from one localized state to another
localized state, from a localized state to an extended state, from an extended
state to a localized state, and from one extended state to another extended
state. At a temperature not too low, any process involving localized state is
activated. The computed mobility of the transitions between localized states
agrees with the observed `hopping mobility'. We suggest that the observed
`drift mobility' originates from the transitions from localized states to
extended states. Analysis of the transition probability from an extended state
to a localized state suggests that there exists a short-lifetime belt of
extended states inside conduction band or valence band. It agrees with the
fact that photoluminescence lifetime decreases with frequency in
a-Si/SiO$_{2}$ quantum well while photoluminescence lifetime is not sensitive
to frequency in c-Si/SiO$_{2}$ structure.

\end{abstract}

\pacs{72.15.Rn, 72.20.Ee, 73.20.Jc, 72.80.Ng}
\keywords{mobility, short-lifetime belt, polaron, electron transfer, reorganization
energy, time evolution of localized state}\maketitle

\section{Introduction}

From the energy spectrum of an amorphous semiconductor\cite{md}, one knows
that there are four types of carrier transitions that may contribute to the
electric conduction: type (1) transition between two localized states; type
(2) transition from a localized state to an extended state; type (3)
transition from an extended state to a localized state and type (4) transition
between two extended states. Type (4) transition is common to both crystalline
and non-crystalline materials. For amorphous semiconductors, type (1)
transition is the main conduction mechanism in a broad temperature range. It
has been investigated in various ways\cite{mul,Over}. Although the thermal
equilibrium population of the extended states is lower than the population of
the localized states, the contribution from the carriers in the extended
states to the electric conduction is still observable for a moderately high
temperature. An investigation of the transitions of type (2), type (3) and
type (4) is necessary for a complete description of carrier dynamics.

In other electronic hopping processes, thermal vibrations of atoms also play
an important role. Electron transfer (in polar solvent and inside large
molecules)\cite{marcu} and polaron diffusion in a molecular
crystal\cite{Hol1,Hol2} are two examples. The transition probability $W_{LL}$
between two sites in a thermally activated process is given by the Marcus
formula%
\begin{equation}
W_{LL}=\nu_{LL}e^{-E_{a}^{LL}/k_{B}T},\text{ \ }E_{a}^{LL}=\frac{\lambda_{LL}%
}{4}(1+\frac{\Delta G_{LL}^{0}}{\lambda_{LL}})^{2} \label{ea}%
\end{equation}
where $\nu_{LL}$ has the dimension of frequency for a specific hopping
process. $E_{a}^{LL}$ is the temperature dependent activation energy.
$\lambda_{LL}$ is the reorganization energy, $\Delta G_{LL}^{0}$ is the energy
difference between the final state and the initial state\cite{marcu}.
Eq.(\ref{ea}) is valid for both electron transfer\cite{marcu} and small
polaron hopping\cite{rei}. The mathematical form of Holstein's work for one
dimensional molecular crystal is quite flexible and can be used to treat three
dimensional materials with slight
modifications\cite{Emin74,Emin75,Emin76,Emin77,Emin91}. The effect of static
disorder may be taken into account by replacing a fixed transfer integral with
a distribution. The static disorder reduces the strength of the
electron-phonon (e-ph) coupling needed to stabilize global small-polaron
fomation\cite{Emin94}. Amorphous semiconductors offer a different regime, in
which the static disorder is so strong that the states in valence and
conduction tails are localized. These localized states interact with the
atomic vibrations.

We will extend Holstein's work\cite{Hol1,Hol2} to amorphous semiconductors. In
Sec.II, we first introduce some notation about localized states, extended
states and electron-phonon interaction. Then the equations of time evolution
for localized states and extended states in presence of atomic vibrations are
derived from a time-dependent Schrodinger equation. The connections with
electron transfer, with Kramers' problem of escape across a barrier and with
small polarons are pointed out.

In Sec.III we study the hopping processes among localized states (LL) in
amorphous solids. Eq.(\ref{ea}) is re-established in high temperature regime
($k_{B}T\gtrsim2.5\hbar\overline{\omega},$ $\overline{\omega}=\int d\omega
D(\omega)\omega/\int d\omega D(\omega)$ is the average phonon frequency,
$D(\omega)$ is the spectral distribution of phonons, the factor 2.5 comes from
the requirement that the error of approximation csch$\frac{\beta\hbar
\overline{\omega}}{2}\thickapprox\frac{2k_{B}T}{\hbar\overline{\omega}}$ is
less than 0.003, cf. the paragraph below Eq.(\ref{dc})). The reorganization
energy $\lambda_{LL}$ is expressed by the phonon spectrum, eigenvectors of the
normal modes and the electron-phonon interaction parameters. The computed
temperature dependence of the mobility of LL transition in a-Si agrees with
that of the observed `hopping' mobility for two regimes T%
$>$%
250K and T%
$<$%
250K (cf. Fig.\ref{fig1} and Fig.\ref{fig2}).

The transition from a localized state to an extended state (LE) induced by the
transfer integral is reported in Sec.IV, it has been suggested as the main
conduction mechanism in amorphous silicon and is called `phonon induced
delocalization'\cite{kik}. Without the dressing effects of the vibrations of
atoms, in certain sense, type (1) transition is similar to the transition
between two bound states in a molecule and to the electron transfer between
two ions, type (2) transition is analogous to the ionization process of an
atom and to the escape across a potential barrier.

The transition from an extended state to a localized state (EL) induced by the
electron-phonon interaction, is presented in Sec.V. In conduction band, the
energy of an extended state is higher than that of a localized state in the
lower tail. For an extended state with high enough energy, EL transition is in
Marcus inverted regime. One expects that there exists a short-lifetime belt of
the extended states inside conduction band or valence band (cf. Fig.\ref{fig3}%
). The states inside this belt favor non-radiative transitions by emitting
several phonons. The conjectured short-lifetime belt in conduction band agrees
the fact that the photoluminescence lifetime of a-Si/SiO$_{2}$ quantum well
decreases with frequency while that of c-Si/SiO$_{2}$ is not sensitive to
frequency. Type (3) transition is similar to the free electron capture process
in an atom and to the capture a particle by a well in a viscous liquid. One
common feature of the type (1), (2) and (3) transitions is that at higher
temperature $k_{B}T\gtrsim2.5\hbar\overline{\omega}$, the probability of any
transition involving localized state takes the form of Eq.(\ref{ea}). We
suggest that the observed drift mobility may originate from type (2)
transitions. Sec.VI is devoted to the transition between two extended states
caused by electron-phonon interaction. Higher order processes and conduction
mechanisms are briefly discussed. Finally we summarize this work and mention
some un-touched problems.

\section{evolution of the states driven by the vibrations}

\subsection{Localized and extended states}

Consider an amorphous sample with $\mathcal{N}$ atoms. Denote the static
positions of the atoms as $\{\mathcal{R}_{\mathbf{n}},$ $\mathbf{n}%
=1,2,\cdots,\mathcal{N}\}$, the displacements of the atoms due to thermal
vibrations as $\{\mathbf{u}^{\mathbf{n}},$ $\mathbf{n}=1,2,\cdots
,\mathcal{N}\}$. To make the formulae compact, we rename the $3\mathcal{N}$
vibrational degrees of freedom $\{u_{x}^{1},u_{y}^{1},u_{z}^{1},\cdots
,u_{x}^{\mathcal{N}},u_{y}^{\mathcal{N}},u_{z}^{\mathcal{N}}\}$ as $\{x_{k},$
$k=1,2,\cdots,3\mathcal{N}\}$ and rename the $3\mathcal{N}$ static position
coordinates $\{\mathcal{R}_{1x},\mathcal{R}_{1y},\mathcal{R}_{1z}%
,\cdots,\mathcal{R}_{\mathcal{N}x},\mathcal{R}_{\mathcal{N}y},\mathcal{R}%
_{\mathcal{N}z}\}$ as $\{X_{j},$ $j=1,2,\cdots,3\mathcal{N}\}$. Consider the
single electron Hamiltonian:%
\begin{equation}
h_{a}=-\frac{\hbar^{2}}{2m}\nabla^{2}+\sum_{\mathbf{n}=1}^{\mathcal{N}%
}U(\mathbf{r}-\mathcal{R}_{\mathbf{n}}) \label{sh}%
\end{equation}
where $U(r-\mathcal{R}_{\mathbf{n}})$ is the potential energy felt by an
electron at $\mathbf{r}$ due to an atom at $\mathcal{R}_{\mathbf{n}}$. In
Eq.(\ref{sh}) only the static disorder of the amorphous sample is taken into
account. $h_{a}$ is a mean field approximation, the existence of $U$ is known
from Hartree-Fock method or density functional theory. $h_{a}$ has two kinds
of eigenstates: localized states $\{\phi_{A_{1}}^{0}\}$%
\begin{equation}
h_{a}\phi_{A_{1}}^{0}(\mathbf{r},\{\mathcal{R}_{\mathbf{n}}\})=E_{A_{1}}%
^{0}(\{\mathcal{R}_{\mathbf{n}}\})\phi_{A_{1}}^{0}(\mathbf{r},\{\mathcal{R}%
_{\mathbf{n}}\}) \label{lo}%
\end{equation}
and extended states $\{\xi_{B_{1}}\}$%
\begin{equation}
h_{a}\xi_{B_{1}}(\mathbf{r},\{\mathcal{R}_{\mathbf{n}}\})=E_{B_{1}%
}(\{\mathcal{R}_{\mathbf{n}}\})\xi_{B_{1}}(\mathbf{r},\{\mathcal{R}%
_{\mathbf{n}}\}) \label{ex}%
\end{equation}
Eigenstates belonging to different eigenvalues are orthogonal to each other.
In conduction band, the energies of the extended states are above those of the
localized states. We use $A_{1},A_{2}\cdots=1,2,\cdots,N_{L}$ to label the
localized states ($N_{L}$ is the total number of the localized states),
$B_{1},B_{2}\cdots$ $=1,2,\cdots,N_{E}$ to label the extended states ($N_{E}$
is the total number of extended states).

When an electron in an extended state is scattered by phonons, the resulting
state is still an extended state. The extended states are not radically
modified by thermal vibrations. The situation for the localized states is
different. The wave function of a localized state $\phi_{A_{1}}^{0}$ is
non-zero only in some finite spatial region $D_{A_{1}}$: only the vibrations
of the atoms inside $D_{A_{1}}$ couple to $\phi_{A_{1}}^{0}$:%
\begin{equation}
\sum_{\mathbf{n}\in D_{A_{1}}}\int d\mathbf{r}\phi_{A_{1}}^{0\ast}%
[\mathbf{u}^{\mathbf{n}}\cdot\nabla_{\mathbf{n}}U(\mathbf{r}-\mathcal{R}%
_{\mathbf{n}})]\phi_{A_{1}}^{0} \label{eph}%
\end{equation}
The `concentrated' wave function makes (\ref{eph}) comparable to or even
larger than the transfer integral between two localized states and the
transfer integral between a localized state and an extended state. The change
in wave function of a localized state must be taken into account in the zeroth
order:%
\begin{equation}
\lbrack-\frac{\hbar^{2}}{2m}\nabla^{2}+\sum_{\mathbf{n}\in D_{A_{1}}%
}U(r-\mathcal{R}_{\mathbf{n}},\mathbf{u}^{\mathbf{n}})]\phi_{A_{1}}%
(\mathbf{r}-\mathcal{R}_{A_{1}},\{x_{p_{A_{1}}}^{A_{1}}\})=E_{A_{1}%
}(\{x_{p_{A_{1}}}^{A_{1}}\})\phi_{A_{1}}(\mathbf{r}-\mathcal{R}_{A_{1}%
},\{x_{p_{A_{1}}}^{A_{1}}\}) \label{0v}%
\end{equation}
where $\mathcal{R}_{A_{1}}$ is a fixed point (could be arbitrarily chosen)
inside $D_{A_{1}}$, $p_{A_{1}}$ is the index of the vibrational degrees of
freedom inside $D_{A_{1}}$. Eq.(\ref{0v}) is a generalization of Eq.(I-3) in
ref.\onlinecite{Hol1}. The effect of the atoms in the neighboring regions of
$D_{A_{1}}$ is neglected. The displacements of the atoms change the overlap
integrals between localized states
\begin{equation}
T_{A_{2}A_{1}}=\int d^{3}r\phi_{A_{2}}^{\ast}(\mathbf{r}-\mathcal{R}_{A_{2}%
},\{x_{p_{A_{2}}}^{A_{2}}\})\phi_{A_{1}}(\mathbf{r}-\mathcal{R}_{A_{1}%
},\{x_{p_{A_{1}}}^{A_{1}}\})=\delta_{A_{2}A_{1}}+S_{A_{2}A_{1}} \label{op1}%
\end{equation}
For two localized states $A_{2}$ and $A_{1}$, $S_{A_{2}A_{1}}$ is a small
quantity if $D_{A_{1}}$ and $D_{A_{2}}$ do not overlap. Using the linear
approximation for e-ph interation\cite{Hol1},
\begin{equation}
E_{A_{1}}(\{x_{p_{A_{1}}}^{A_{1}}\})=E_{A_{1}}^{0}-\sum_{p_{A_{1}}\mathbf{\in
}D_{A_{1}}}d_{p_{A_{1}}}x_{p_{A_{1}}}^{A_{1}},\text{ \ \ \ \ }d_{p_{A_{1}}%
}=\int d\mathbf{r}\phi_{A_{1}}^{0}(\mathbf{r},\{\mathcal{R}_{\mathbf{n}%
}\})\frac{\partial U}{\partial X_{p_{A_{1}}}}\phi_{A_{1}}^{0}(\mathbf{r}%
,\{\mathcal{R}_{\mathbf{n}}\}) \label{lin1}%
\end{equation}
Eq.(\ref{lin1}) is the first order correction of electronic energy due to the
e-ph interaction. The most localized states can be described by $\xi
^{-3/2}e^{-r/\xi},$ where $\xi$ is localization length of a localized state,
$r$ is the distance between electron and some representative point inside
$D_{A_{1}}$. $d_{p_{A_{1}}}\thicksim\frac{Z^{\ast}e^{2}}{4\pi\epsilon_{0}%
\xi^{2}},$ $\epsilon_{0}$ is the permittivity of vacuum, $Z^{\ast}$ is the
effective nuclear charge of an atom felt by a conduction electron. For
electrons in a metal, researchers usually focus on how an electron in an
extended state is scattered into another extended state by the e-ph
interaction rather than the correction to energy. Unlike $\phi_{A_{1}}^{0}$,
$\phi_{A_{1}}$ includes the effect of vibrations of the atoms and is no longer
orthogonal to any extended states of (\ref{sh}),%
\begin{equation}
\int d^{3}r\xi_{B_{2}}^{\ast}(\mathbf{r},\{\mathcal{R}_{\mathbf{n}}%
\})\phi_{A_{1}}(\mathbf{r}-\mathcal{R}_{A_{1}},\{x_{p_{A_{1}}}^{A_{1}%
}\})=Y_{B_{2}A_{1}} \label{ope}%
\end{equation}

\subsection{Single-electron approximation}

For definiteness, we consider the electrons in the conduction band of an
amorphous semiconductor. For carriers in the mid-gap states and the holes in
valence band, we need only slightly modify the notation. In intrinsic and
lightly doped n-type semiconductors, the number of the electrons is much
smaller than the number of the localized states. The correlation between
electrons in a hopping process and the screen effect caused by these electrons
can be neglected. Essentially we have a single particle problem: one electron
moves in many empty localized states and extended states in the conduction band.

Consider one electron moving in an amorphous solid with $\mathcal{N}$ atoms,
the total Hamiltonian of the system is%
\begin{equation}
H=\frac{-\hbar^{2}}{2m}\nabla^{2}+\sum_{\mathbf{n}=1}^{\mathcal{N}%
}U(\mathbf{r},\mathbf{W}_{\mathbf{n}})+\sum_{\mathbf{n}}\frac{-\hbar^{2}%
}{2M_{\mathbf{n}}}\nabla_{\mathbf{n}}^{2}+\sum_{\mathbf{n,n}^{\prime
}\mathbf{(n\neq n}^{\prime})}V(\mathbf{W}_{\mathbf{n}},\mathbf{W}%
_{\mathbf{n}^{\prime}}) \label{sing}%
\end{equation}
where $\mathbf{n}$ denote the site of network, $\mathbf{W}_{\mathbf{n}}$ is
the instantaneous position vector of the nucleus at site $\mathbf{n}$,
$U(\mathbf{r},\mathbf{W}_{\mathbf{n}})$ is the interaction between the
electron and the nucleus at the $\mathbf{n}^{\text{th}}$ site. $V(\mathbf{W}%
_{\mathbf{n}},\mathbf{W}_{\mathbf{n}^{\prime}})$ is the total effective
interaction of the nucleus at site $\mathbf{n}^{\prime}$ and the nucleus at
site $\mathbf{n},$ which including both the Coulomb repulsion between them and
the induced attraction by the electrons. The total wave function
$\psi(\mathbf{r},\{\mathbf{W}_{\mathbf{n}}\};t)$ of the system is a function
of all degrees of freedom, its time evolution is determined by%
\begin{equation}
i\hbar\frac{\partial}{\partial t}\psi(\mathbf{r},\{\mathbf{W}_{\mathbf{n}%
}\};t)=H\psi(\mathbf{r},\{\mathbf{W}_{\mathbf{n}}\};t)\text{.} \label{tim}%
\end{equation}
For temperature well below melting point, the system executes small harmonic
oscillations. The motion of nuclei $\mathbf{W}_{\mathbf{n}}=\mathcal{R}%
_{\mathbf{n}}+\mathbf{u}^{\mathbf{n}}$ can be viewed as vibrations of the
atoms $\{\mathbf{u}^{\mathbf{n}}\}$ around their equilibrium positions
$\{\mathcal{R}_{\mathbf{n}}\}$. $\psi(\mathbf{r},\{\mathbf{W}_{\mathbf{n}%
}\};t)$ is changed into $\psi(\mathbf{r},\{\mathbf{u}^{\mathbf{n}}\};t)$, a
function of displacements of the atoms. $H$ is simplified as%
\begin{equation}
H_{1}=h_{e}+h_{v} \label{has}%
\end{equation}
where%
\begin{equation}
h_{e}=\frac{-\hbar^{2}}{2m}\nabla^{2}+\sum_{\mathbf{n}=1}^{\mathcal{N}%
}U(\mathbf{r},\mathcal{R}_{\mathbf{n}},\mathbf{u}^{\mathbf{n}}),\text{
\ \ }h_{v}=\sum_{j}-\frac{\hbar^{2}}{2M_{j}}\nabla_{j}^{2}+\frac{1}{2}%
\sum_{jk}k_{jk}x_{j}x_{k} \label{hes}%
\end{equation}
$h_{e}$ is the single electron Hamiltonian including the vibrations. $h_{a}$
in Eq.(\ref{sh}) and the Hamiltonian in Eq.(\ref{0v}) are two different
approximations of $h_{e}$. $h_{v}$ is the vibrational Hamiltonian, $(k_{jk})$
is the matrix of force constants. The evolution of the total wave function
$\psi(\mathbf{r},\{\mathbf{u}^{\mathbf{n}}\};t)$ of the system of
\textquotedblleft one electron + many nuclei\textquotedblright\ is given by
Schrodinger equation%
\begin{equation}
i\hbar\frac{\partial\psi(\mathbf{r},\{\mathbf{u}^{\mathbf{n}}\};t)}{\partial
t}=H_{1}\psi(\mathbf{r},\{\mathbf{u}^{\mathbf{n}}\};t) \label{fs1}%
\end{equation}

\subsection{Evolution equations}

The Hilbert space of $h_{e}$ is spanned by the localized states and the
extended states. The total wave function $\psi(\mathbf{r},\{\mathbf{u}%
^{\mathbf{n}}\};t)$ of the system of \textquotedblleft one electron + many
nuclei\textquotedblright\ can be expanded as%

\begin{equation}
\psi(\mathbf{r},x_{1},\cdots,x_{3\mathcal{N}};t)=\sum_{A_{1}}a_{A_{1}}%
(x_{1},\cdots,x_{3\mathcal{N}};t)\phi_{A_{1}}+\sum_{B_{1}}b_{B_{1}}%
(x_{1},\cdots,x_{3\mathcal{N}};t)\xi_{B_{1}} \label{fw1}%
\end{equation}
where $a_{A_{1}}$ is the probability amplitude at moment $t$ that the electron
is in localized state $A_{1}$ while the displacements of the nuclei are
$\{x_{j},$ $j=1,2,\cdots3\mathcal{N}\}$, $b_{B_{1}}$ is the amplitude at
moment $t$ that the electron is in extended state $B_{1}$ while the
displacements of the nuclei are $\{x_{j},$ $j=1,2,\cdots3\mathcal{N}\}$. The
first sum runs over all the localized states, the second sum runs over all the
extended states. Substitute Eq.(\ref{fw1}) into Eq.(\ref{fs1}), we have%
\begin{equation}
\sum_{A_{1}}i\hbar\frac{\partial a_{A_{1}}}{\partial t}\phi_{A_{1}}%
+\sum_{B_{1}}i\hbar\frac{\partial b_{B_{1}}}{\partial t}\xi_{B_{1}}%
=\sum_{A_{1}}a_{A_{1}}E_{A_{1}}\phi_{A_{1}}+\sum_{A_{1}}a_{A_{1}}%
\sum_{\mathbf{p}\notin D_{A_{1}}}U(r-\mathcal{R}_{\mathbf{p}},\mathbf{u}%
_{\mathbf{p}})\phi_{A_{1}} \label{se}%
\end{equation}%
\[
+\sum_{A_{1}}\phi_{A_{1}}h_{v}a_{A_{1}}+\sum_{A_{1}}\sum_{j}-\frac{\hbar^{2}%
}{M_{j}}(\nabla_{j}a_{A_{1}})(\nabla_{j}\phi_{A_{1}})+\sum_{A_{1}}\sum
_{j}-\frac{\hbar^{2}}{2M_{j}}a_{A_{1}}\nabla_{j}^{2}\phi_{A_{1}}%
\]%
\[
+\sum_{B_{1}}b_{B_{1}}E_{B_{1}}\xi_{B_{1}}+\sum_{B_{1}}b_{B_{1}}\sum_{j}%
x_{j}\frac{\partial U}{\partial X_{j}}\xi_{B_{1}}+\sum_{B_{1}}\xi_{B_{1}}%
\sum_{j}h_{v}b_{B_{1}}%
\]%
\[
+\sum_{B_{1}}\sum_{j}-\frac{\hbar^{2}}{M_{j}}(\nabla_{j}b_{B_{1}})(\nabla
_{j}\xi_{B_{1}})+\sum_{B_{1}}\sum_{j}-\frac{\hbar^{2}}{2M_{j}}b_{B_{1}}%
\nabla_{j}^{2}\xi_{B_{1}}%
\]
In extended states $\{\xi_{B_{1}}\}$, we neglected the dependence of on the
vibrational displacements, the last two terms disappear.

Differences between the localized states and the extended states are reflected
in Eq.(\ref{se}). For a localized state, we have separated%
\begin{equation}
\sum_{\mathbf{n}}U(r-\mathcal{R}_{\mathbf{n}},\mathbf{u}^{\mathbf{n}}%
)=\sum_{\mathbf{n}\in D_{A_{1}}}U(r-\mathcal{R}_{\mathbf{n}},\mathbf{u}%
^{\mathbf{n}})+\sum_{\mathbf{p}\notin D_{A_{1}}}U(r-\mathcal{R}_{\mathbf{p}%
},\mathbf{u}^{\mathbf{p}}) \label{sl}%
\end{equation}
the second term leads to the transfer integral which causes transition among
states. The wave function of localized state $A_{1}$ is confined in $D_{A_{1}%
}$. For a nucleus outside $D_{A_{1}}$, its effect on $A_{1}$ dies away with
the distance between the nucleus and $D_{A_{1}}$. Eq.(\ref{sl}) is a
generalization of Holstein's treatment\cite{Hol1} to a localized state which
occupies several sites. While for extended states we have resolved%
\begin{equation}
\sum_{\mathbf{n}}U(r-\mathcal{R}_{\mathbf{n}},\mathbf{u}^{\mathbf{n}}%
)=\sum_{\mathbf{n}=1}^{\mathcal{N}}U(\mathbf{r}-\mathcal{R}_{\mathbf{n}}%
)+\sum_{j=1}^{3\mathcal{N}}x_{j}\frac{\partial U}{\partial X_{j}} \label{es}%
\end{equation}
the second term is the electron-phonon interaction. An extended state spreads
over whole sample, it feels the vibrations of all the atoms. Eq.(\ref{es}) is
similar to the usual treatment of e-ph interaction in a crystal.

The first sum in Eq.(\ref{sl}) only includes the atoms in the region where the
wave function of localized state $A_{1}$ is nonzero. If a localized state is
close to mobility edge, it extends to a very large spatial region. For such a
state, Eq.(\ref{sl}) is not very different from Eq.(\ref{es}).

By applying $\int d^{3}r\phi_{A_{2}}^{\ast}$ to both sides of Eq.(\ref{se}),
one obtains:%
\begin{equation}
\sum_{A_{1}}T_{A_{2}A_{1}}\{i\hbar\frac{\partial}{\partial t}-E_{A_{1}}%
-h_{v}\}a_{A_{1}}+\sum_{B_{1}}Y_{A_{2}B_{1}}^{\ast}\{i\hbar\frac{\partial
}{\partial t}-E_{B_{1}}-h_{v}\}b_{B_{1}}= \label{se1}%
\end{equation}%
\[
\sum_{A_{1}}a_{A_{1}}\int d^{3}r\phi_{A_{2}}^{\ast}\sum_{\mathbf{p}\notin
D_{A_{1}}}U(r-\mathcal{R}_{\mathbf{p}},\mathbf{u}_{\mathbf{p}})\phi_{A_{1}%
}+\sum_{B_{1}}b_{B_{1}}\int d^{3}r\phi_{A_{2}}^{\ast}\sum_{j}x_{j}%
\frac{\partial U}{\partial X_{j}}\xi_{B_{1}}%
\]%
\[
+\sum_{A_{1}}\sum_{j}-\frac{\hbar^{2}}{M_{j}}(\nabla_{j}a_{A_{1}})\int
d^{3}r\phi_{A_{2}}^{\ast}(\nabla_{j}\phi_{A_{1}})+\sum_{A_{1}}\sum_{j}%
-\frac{\hbar^{2}}{2M_{j}}a_{A_{1}}\int d^{3}r\phi_{A_{2}}^{\ast}\nabla_{j}%
^{2}\phi_{A_{1}}%
\]
Application of $\int d^{3}r\xi_{B_{2}}^{\ast}$ to both sides of Eq.(\ref{se})
yields:
\begin{equation}
\sum_{A_{1}}Y_{B_{2}A_{1}}\{i\hbar\frac{\partial}{\partial t}-E_{A_{1}}%
-h_{v}\}a_{A_{1}}+\{i\hbar\frac{\partial}{\partial t}-E_{B_{2}}-h_{v}%
\}b_{B_{2}} \label{se3}%
\end{equation}%
\[
=\sum_{A_{1}}a_{A_{1}}\sum_{\mathbf{p}\notin D_{A_{1}}}\int d^{3}r\xi_{B_{2}%
}^{\ast}U(r-\mathcal{R}_{\mathbf{p}},\mathbf{u}^{\mathbf{p}})\phi_{A_{1}}%
+\sum_{B_{1}}b_{B_{1}}\sum_{j}x_{j}\int d^{3}r\xi_{B_{2}}^{\ast}\frac{\partial
U}{\partial X_{j}}\xi_{B_{1}}%
\]%
\[
+\sum_{A_{1}}\sum_{j}-\frac{\hbar^{2}}{M_{j}}(\nabla_{j}a_{A_{1}})\int
d^{3}r\xi_{B_{2}}^{\ast}(\nabla_{j}\phi_{A_{1}})+\sum_{A_{1}}\sum_{j}%
-\frac{\hbar^{2}}{2M_{j}}a_{A_{1}}\int d^{3}r\xi_{B_{2}}^{\ast}\nabla_{j}%
^{2}\phi_{A_{1}}%
\]
The coupled equations Eq.(\ref{se1}) and Eq.(\ref{se3}) describe the time
evolution of the states under the influence of the vibrations: the transition
between two localized states (LL), the transition between two extended states
(EE), the transition from an extended state to a localized state (EL) and the
transition from a localized state to an extended state (LE). In other words,
for a given initial state, i.e. $\{a_{A_{1}}(x_{1},\cdots,x_{3\mathcal{N}%
};t=0),A_{1}=1,2,\cdots,N_{L}\}$ and $\{b_{B_{1}}(x_{1},\cdots,x_{3\mathcal{N}%
};t=0),B_{1}=1,2,\cdots,N_{E}\}$, at any subsequent moment $t>0$ the state of
the system is completely determined (i.e $\{a_{A_{1}}(x_{1},\cdots
,x_{3\mathcal{N}};t),A_{1}=1,2,\cdots,N_{L}\}$ and $\{b_{B_{1}}(x_{1}%
,\cdots,x_{3\mathcal{N}};t),B_{1}=1,2,\cdots,N_{E}\}$) by Eqs. (\ref{se1}) and
(\ref{se3}). The number of vibrational degrees of freedom is macroscopic, so
that $\{a_{A_{1}}(t=0),A_{1}=1,2,\cdots,N_{L}\}$ and $\{b_{B_{1}}%
(t=0,B_{1}=1,2,\cdots,N_{E}\}$ cannot be assigned precisely. A description
based on density matrix is more appropriate.

Let us estimate the order of magnitude of the last two terms in RHS of
Eq.(\ref{se1}) or Eq.(\ref{se3}). From the 1$^{\text{st}}$ order perturbation
correction about the nuclear displacements to the localized wave function, we
have%
\begin{equation}
\nabla_{j}\phi_{A_{1}}=\sum_{C}{}^{\prime}\frac{\langle C|\sum_{j\in A_{1}%
}\frac{\partial U}{\partial X_{j}}|A_{1}\rangle}{E_{A_{1}}-E_{C}}%
|C\rangle\thicksim\frac{U}{\hbar v_{e}} \label{dao}%
\end{equation}
where $v_{e}$ is a typical velocity of electron, $C$ is eigenvalue $A$ or $B$
of $h_{a}$. To reach last step, we notice the derivative of $U(\mathbf{r}%
-\mathcal{R}_{j})$ respect to suitable coordinate component of electron equals
$-\frac{\partial U}{\partial X_{j}}$. Thus $\frac{\partial U}{\partial X_{j}%
}\thicksim p_{e}U/\hbar$, $p_{e}$ is momentum of the electron, and
$v_{e}=\frac{\Delta E}{\Delta p_{e}}$. A typical term in the 2$^{\text{nd}}$
sum from last in RHS of Eq.(\ref{se1}) becomes
\begin{equation}
\frac{\hbar^{2}}{M_{j}}\nabla_{j}a_{A_{1}}\cdot\nabla_{j}\phi_{A_{1}}\text{
}\thicksim\frac{v_{n}}{v_{e}}U\text{ } \label{dao2}%
\end{equation}
where $v_{n}\thicksim M_{j}^{-1}\hbar\nabla_{j}$ is typical velocity of a
nucleus. A typical term in the last sum in RHS of Eq.(\ref{se1}) can be
estimated from 2nd order correction to $\phi_{A_{1}}$ about the nuclear
displacements \
\begin{equation}
\frac{\hbar^{2}}{2M_{j}}\nabla_{j}^{2}\phi_{A_{1}}\thicksim\frac{U^{2}}%
{Mv_{e}^{2}}\thicksim\frac{m}{M}U \label{dao1}%
\end{equation}
In the last $\thicksim$, we applied the virial theorem $mv_{e}^{2}\thicksim
U$, $m$ is the mass of electron, $U$ is the interaction potential energy
between electron and some nucleus. Clearly%
\begin{equation}
x_{j}\frac{\partial U}{\partial X_{j}}\thicksim\frac{x}{d}U \label{1eph}%
\end{equation}
where $x$ is a typical vibrational amplitude of atoms, $d$ is the distance
between two nearest neighbor atoms. Notice $v_{n}/v_{e}\thicksim10^{-3}$,
$m/M\thicksim10^{-4}$ and $x/d\thicksim10^{-2}-10^{-1}$, combining Eqs.
(\ref{dao2}), (\ref{dao1}) and (\ref{1eph}), we find
\begin{equation}
\sum_{j}\frac{\hbar^{2}}{M_{j}}(\nabla_{j}a_{A_{1}})(\nabla_{j}\phi_{A_{1}%
})\text{ or }\sum_{j}\frac{\hbar^{2}}{2M_{j}}\nabla_{j}^{2}\phi_{A_{1}}%
<<\sum_{\mathbf{p}\notin D_{A_{1}}}U(r-\mathcal{R}_{\mathbf{p}},\mathbf{u}%
^{\mathbf{p}})\phi_{A_{1}}\text{ or }\sum_{j}x_{j}\frac{\partial U}{\partial
X_{j}}\xi_{B_{1}} \label{order}%
\end{equation}
The 3rd term and the 4th term in the RHS of both Eq.(\ref{se1}) and
Eq.(\ref{se3}) can be ignored. Eq.(\ref{se1}) is reduced to%
\begin{equation}
\sum_{A_{1}}T_{A_{2}A_{1}}\{i\hbar\frac{\partial}{\partial t}-E_{A_{1}}%
-h_{v}\}a_{A_{1}}+\sum_{B_{1}}Y_{A_{2}B_{1}}^{\ast}\{i\hbar\frac{\partial
}{\partial t}-E_{B_{1}}-h_{v}\}b_{B_{1}}= \label{se2}%
\end{equation}%
\[
\sum_{A_{1}}a_{A_{1}}\int d^{3}r\phi_{A_{2}}^{\ast}\sum_{\mathbf{p}\notin
D_{A_{1}}}U(r-\mathcal{R}_{\mathbf{p}},\mathbf{u}^{\mathbf{p}})\phi_{A_{1}%
}+\sum_{B_{1}}b_{B_{1}}\int d^{3}r\phi_{A_{2}}^{\ast}\sum_{j}x_{j}%
\frac{\partial U}{\partial X_{j}}\xi_{B_{1}}%
\]
In the right hand side (RHS), each term in the first sum is a transition
between two localized states mediated by a transfer integral. Each term in the
second sum is a transition from an extended state to a localized state caused
by the electron-phonon interaction. Similarly Eq.(\ref{se3}) is reduced to
\begin{equation}
\sum_{A_{1}}Y_{B_{2}A_{1}}\{i\hbar\frac{\partial}{\partial t}-E_{A_{1}}%
-h_{v}\}a_{A_{1}}+\{i\hbar\frac{\partial}{\partial t}-E_{B_{2}}-h_{v}%
\}b_{B_{2}}= \label{se4}%
\end{equation}%
\[
\sum_{A_{1}}a_{A_{1}}\sum_{\mathbf{p}\notin D_{A_{1}}}\int d^{3}r\xi_{B_{2}%
}^{\ast}U(r-\mathcal{R}_{\mathbf{p}},\mathbf{u}^{\mathbf{p}})\phi_{A_{1}}%
+\sum_{B_{1}}b_{B_{1}}\sum_{j}x_{j}\int d^{3}r\xi_{B_{2}}^{\ast}\frac{\partial
U}{\partial X_{j}}\xi_{B_{1}}%
\]
In the RHS, each term in the first sum is a transition from a localized state
to an extended state induced by a transfer integral. Each term in the second
sum is a transition between two extended states caused by electron-phonon interaction.

Eqs.(\ref{se2}) and (\ref{se4}) are completely general. To simplify them we
need two connected technical assumptions (i) $Y_{B_{2}A_{1}}\thicksim
\frac{N_{A_{1}}}{\mathcal{N}}<<1$ and (ii) $S_{A_{2}A_{1}}<<1$. Obviously they
are not true in general. For localized states which are close to mobility
edge, they spread in many distorted spatial regions. The overlap integral
$Y_{B_{2}A_{1}}$ between them and an extended state is not small (when e-ph
interaction appears, they are no longer eigenstates of $h_{a}$). Assumption
(i) means that we do not consider the localized states very close to mobility
edge and consider only the most localized states. The contributions to
conductivity from near-edge localized states are ignored. For the most
localized states (the most low-lying ones in conduction band), which spread
over only several bond lengths at most, condition (i) is satisfied. Condition
(ii) is satisfied for two localized states which do not overlap. It means we
exclude the indirect contribution to conductivity from the transitions between
two localized states with overlapping spatial regions. For two eigenstates
$A_{2}$ and $A_{1}$ of static hamiltonian $h_{a}$, the overlap integral
between two states $S_{A_{2}A_{1}}=0.$ In addition, if the distance between
two localized states is larger than one bond length, the overlap integral can
be neglected even one takes into account electron-phonon interaction (e-ph).
When e-ph interaction is taken into account, $A_{2}$ and $A_{1}$ are no longer
eigenstates of $h_{a}$, $S_{A_{2}A_{1}}$ is not negligible when the spatial
regions of two localized states overlap. For a localized state which spreads
in several distorted regions\cite{dong,lud}, its wave function is not single
exponential decay function which is only suitable for the most localized
states\cite{big}. The overlap integrals $Y_{B_{2}A_{1}}$ and $S_{A_{2}A_{1}}$
involving such a localized state are oscillatory. To make a semi-classical
estimation, one needs detailed information of wavefunction which could be
obtained through a WKB-like exponential transform. We wish to remove two
assumptions in a later communication\cite{wkb}.

The transition between two localized states is significant only when the
distance $R_{12}$ between the two is not very large. For two localized states
$A_{2}$ and $A_{1}$, $S_{A_{2}A_{1}}<<1$ if $D_{A_{1}}$ and $D_{A_{2}}$ do not
overlap. The terms multiplied by $S_{A_{2}A_{1}}$ can be neglected for
localized states which their spatial regions do not overlap. What is more, the
transfer integral is important only when the atoms $\mathbf{p}\notin D_{A_{1}%
}$ fall into $D_{A_{2}}$ or $A_{1}=A_{2}$,%

\begin{equation}
\sum_{A_{1}}a_{A_{1}}\int d^{3}r\phi_{A_{2}}^{\ast}\sum_{\mathbf{p}\notin
D_{A_{1}}}U(r-\mathcal{R}_{\mathbf{p}},\mathbf{u}^{\mathbf{p}})\phi_{A_{1}%
}\thickapprox\sum_{A_{1}}a_{A_{1}}J_{A_{2}A_{1}}+W_{A_{2}}a_{A_{2}%
}\thickapprox\sum_{A_{1}}a_{A_{1}}J_{A_{2}A_{1}} \label{tra}%
\end{equation}
where%
\begin{equation}
J_{A_{2}A_{1}}=\int d^{3}r\phi_{A_{2}}^{\ast}\sum_{\mathbf{p}\in D_{A_{2}}%
}U(r-\mathcal{R}_{\mathbf{p}},\mathbf{u}^{\mathbf{p}})\phi_{A_{1}},\text{
\ \ \ }W_{A_{2}}=\int d^{3}r|\phi_{A_{2}}|^{2}\sum_{\mathbf{p}\notin D_{A_{2}%
}}U(r-\mathcal{R}_{\mathbf{p}},\mathbf{u}^{\mathbf{p}}) \label{tra1}%
\end{equation}
\ Here, the $W_{A_{2}}$ term only affects the self energy of a localized state
through $a_{A_{2}}$. Eqs.(\ref{tra}) and (\ref{tra1}) are a generalization of
Eqs.(I-14) to (I-16) of ref.\onlinecite{Hol1}. Comparing with $E_{A_{1}}$ and
with $h_{v}$, $W_{A_{2}}$ can be neglected. $J_{A_{2}A_{1}}$ causes the
transition from $A_{1}$ to $A_{2}$, it comes from the attraction on the
electron by the atoms in $D_{A_{2}}$. For those most localized states, the
wave functions take form of $\phi_{A_{1}}\thicksim e^{-|\mathbf{r}%
-\mathcal{R}_{A_{1}}|/\xi_{1}}$. $J_{A_{2}A_{1}}$ is estimated to be\cite{pau}
$-\frac{N_{A_{2}}Z^{\ast}e^{2}}{4\pi\epsilon_{0}\varepsilon_{s}\xi}%
(1+\frac{R_{12}}{\xi})e^{-R_{12}/\xi}$, where average localization length
$\xi$ is defined by $2\xi^{-1}=\xi_{1}^{-1}+\xi_{2}^{-1}$. $R_{12}$ is the
average distance between two localized states, $\varepsilon_{s}$ is static
dielectric function, $Z^{\ast}$ is the effective nuclear charge of atom,
$N_{A_{2}}$ is the number of atoms inside region $D_{A_{2}}$.

Similarly $J_{A_{1}A_{2}}$ causes the transition from $A_{2}$ to $A_{1}$, it
comes from the attraction on the electron by the atoms in $D_{A_{1}}$. For two
localized states $A_{1}$ and $A_{2}$ in different regions, no simple relation
exists between $J_{A_{2}A_{1}}$ and $J_{A_{1}A_{2}}$. This is in contrast with
the situation $J_{A_{2}A_{1}}=(J_{A_{1}A_{2}})^{\ast}$ of small polarons in
crystal where translational invariance exists \cite{Hol2}. Later we neglect
the dependence of $J_{A_{2}A_{1}}$ on the displacements $\{x\}$ of the atoms
and consider $J_{A_{2}A_{1}}$ as a function of the distance $R_{12}$ between
two localized states, localization length $\xi_{1}$ of state $A_{1}$ and
localization length $\xi_{2}$ of state $A_{2}$.

Eqs.(\ref{se2}) and (\ref{se4}) are then reduced to%
\begin{equation}
\{i\hbar\frac{\partial}{\partial t}-E_{A_{2}}-h_{v}\}a_{A_{2}}+\sum_{B_{1}%
}Y_{A_{2}B_{1}}^{\ast}\{i\hbar\frac{\partial}{\partial t}-E_{B_{1}}%
-h_{v}\}b_{B_{1}}=\sum_{A_{1}}J_{A_{2}A_{1}}a_{A_{1}}+\sum_{B_{1}}%
K_{A_{2}B_{1}}^{\prime}b_{B_{1}} \label{s1}%
\end{equation}
and
\begin{equation}
\sum_{A_{1}}Y_{B_{2}A_{1}}\{i\hbar\frac{\partial}{\partial t}-E_{A_{1}}%
-h_{v}\}a_{A_{1}}+\{i\hbar\frac{\partial}{\partial t}-E_{B_{2}}-h_{v}%
\}b_{B_{2}}=\sum_{A_{1}}J^{\prime}{}_{B_{2}A_{1}}a_{A_{1}}+\sum_{B_{1}%
}K_{B_{2}B_{1}}b_{B_{1}} \label{s2}%
\end{equation}
where%

\begin{equation}
K_{A_{2}B_{1}}^{\prime}=\sum_{j}x_{j}\int d^{3}r\phi_{A_{2}}^{\ast}%
\frac{\partial U}{\partial X_{j}}\xi_{B_{1}} \label{ept}%
\end{equation}
is a linear function of atomic displacements $x_{j}$. It causes type (3)
transition from an extended state to a localized state. If we approximate
extended state as plane wave $\xi_{B_{1}}\thicksim e^{ik_{B_{1}}r}$,
$K_{A_{2}B_{1}}^{\prime}\thicksim\frac{Z^{\ast}e^{2}u}{4\pi\epsilon
_{0}\varepsilon_{s}\xi_{A_{2}}^{2}}(1-ik_{B_{1}}\xi_{A_{2}})^{-1}$, where
$u\thicksim\sqrt{\frac{k_{B}T}{M\omega^{2}}}$ or $\sqrt{\frac{\hbar}{M\omega}%
}$ is typical amplitude of vibration at high or low temperature. So that
$K_{A_{2}B_{1}}^{\prime}u/J_{A_{2}A_{1}}\thicksim e^{R_{12}/\xi}u/\xi$. The
distance between two nearest localized states is about several \AA \ in a-Si,
$K_{A_{2}B_{1}}^{\prime}u$ is several times smaller than $J_{A_{2}A_{1}}$.

From Eqs.(\ref{s1}) and (\ref{s2}), type (2) transition from a localized state
in region $D_{A_{1}}$ to an extended state is caused by the transfer integral%
\begin{equation}
J^{\prime}{}_{B_{2}A_{1}}=\sum_{\mathbf{p}\notin D_{A_{1}}}\int d^{3}%
r\xi_{B_{2}}^{\ast}U(r-\mathcal{R}_{\mathbf{p}},\mathbf{u}^{\mathbf{p}}%
)\phi_{A_{1}} \label{tra2}%
\end{equation}
not by the electron-phonon interaction $K_{A_{2}B_{1}}^{\prime}$ (although we
will see the process does involve several phonons in high temperature regime
in Sec.IV, it consists with the intuitive picture of `phonon induced
delocalization'\cite{kik}). $J^{\prime}{}_{B_{2}A_{1}}$ does not involve
atomic displacements explicitly. Later we neglect the dependence of
$J^{\prime}{}_{B_{2}A_{1}}$ on the displacements of atoms and only view
$J^{\prime}{}_{B_{2}A_{1}}$ as function of $\xi_{A_{1}}$ only. If we
approximate extended state as plane wave $\xi_{B_{2}}^{\ast}\thicksim
e^{-ik_{B_{2}}r}$, $J^{\prime}{}_{B_{2}A_{1}}\thicksim\frac{Z^{\ast}e^{2}%
}{4\pi\epsilon_{0}\varepsilon_{s}\xi_{A_{1}}}(1+ik_{B_{2}}\xi_{A_{1}})^{-2}$.
$J^{\prime}{}_{B_{2}A_{1}}$ is in the same order of magnitude as
$J_{A_{2}A_{1}}$. According to Eqs.(\ref{s1}) and (\ref{s2}), $J^{\prime}%
{}_{B_{2}A_{1}}$ does not create transitions from an extended state to a
localized state. The asymmetries in Eq.(\ref{ept}) and Eq.(\ref{tra2}) come
from the different separations Eq.(\ref{sl}) and Eq.(\ref{es}) of the single
particle potential energy for localized states and extended states. One should
not confuse with the usual symmetry between transition probabilities for
forward process and backward process computed by the first order perturbation
theory, where two processes are coupled by the \textit{same} interaction.

Type (4) transition between two extended states $B_{1}$ and $B_{2}$ is caused
by electron-phonon interaction:%
\begin{equation}
K_{B_{2}B_{1}}=\sum_{j}x_{j}\int d^{3}r\xi_{B_{2}}^{\ast}\frac{\partial
U}{\partial X_{j}}\xi_{B_{1}},\text{ \ \ \ }K_{B_{2}B_{1}}=(K_{B_{1}B_{2}%
})^{\ast} \label{ep0}%
\end{equation}
It is almost the same as the usual scattering between two Bloch states in a
crystal by the electron-phonon interaction. If we approximate extended states
$\xi_{B_{1}}$ and $\xi_{B_{2}}$ by plane waves with wave vector $k_{1}$ and
$k_{2}$, $K_{B_{2}B_{1}}\thicksim\frac{Z^{\ast}e^{2}u\kappa^{3}}{4\pi
\epsilon_{0}\varepsilon_{s}[\kappa+i(k_{2}-k_{1})]}$, where $\kappa
\thicksim\frac{e^{2}}{\epsilon_{0}}\frac{\partial n}{\partial\mu}$ is the
Thomas-Fermi screening wave vector. In lightly doped or intrinsic
semiconductor, $\kappa$ is hundreds even thousands times smaller than $1/a,$
$a$ is bond length. Since for most localized state, localization length $\xi$
is several times $a$, $K_{B_{2}B_{1}}\thicksim(\kappa\xi)^{2}K_{A_{2}B_{1}%
}^{\prime}$, is much weaker than three other coupling constants. Contrast with
type (1), type (2) and type (3) transitions, the transition probability of the
transition $B_{1}\rightarrow B_{2}$ equals to $B_{2}\rightarrow B_{1}$: two
processes are coupled by the \textit{same} interaction as illustrated in
Eq.(\ref{ep0}). Eq.(\ref{s1}) and Eq.(\ref{s2}) correspond to the generalized
master equation for the reduced density matrix of an electron in a phonon
bath\cite{mas}.

In the conduction band of an amorphous solid, the energies of extended states
are higher than the energies of \ the localized states close to the bottom of
the band. If a localized state is not very close to the mobility edge, its
localization length $\xi$ is small. The overlap between it and an extended
state $Y_{B_{2}A_{1}}\thicksim\frac{N_{A_{1}}}{\mathcal{N}}$ may be neglected.
Eq.(\ref{s1}) and Eq.(\ref{s2}) read as%

\begin{equation}
\{i\hbar\frac{\partial}{\partial t}-E_{A_{2}}-h_{v}\}a_{A_{2}}=\sum_{A_{1}%
}J_{A_{2}A_{1}}a_{A_{1}}+\sum_{B_{1}}K_{A_{2}B_{1}}^{\prime}b_{B_{1}}
\label{s11}%
\end{equation}
and
\begin{equation}
\{i\hbar\frac{\partial}{\partial t}-E_{B_{2}}-h_{v}\}b_{B_{2}}=\sum_{A_{1}%
}J^{\prime}{}_{B_{2}A_{1}}a_{A_{1}}+\sum_{B_{1}}K_{B_{2}B_{1}}b_{B_{1}}
\label{s22}%
\end{equation}
If a localized state is close to the mobility edge, the overlaps between it
and the extended states are not necessarily small. Eqs. (\ref{s11}) and
(\ref{s22}) could be used generically with the restrictions (i) and (ii).

In Eq.(\ref{s11}) and Eq.(\ref{s22}), the e-ph interaction for localized state
and extended state has been treated differently. The reason is as following.
Let us consider the potential energy of atoms%
\begin{equation}
\frac{1}{2}\sum_{jk}k_{jk}x_{j}x_{k}-\sum_{p}g_{p}x_{p},\text{ \ \ \ }%
g_{p}=-\frac{\partial U}{\partial X_{p}} \label{pot}%
\end{equation}
the first term is the mutual interactions among atoms. The second term, the
e-ph interaction, acts like an external field. The first member of
Eq.(\ref{pot}) can be written as:%
\begin{equation}
\frac{1}{2}\sum_{jk}k_{jk}(x_{j}-x_{j}^{0})(x_{k}-x_{k}^{0})-\frac{1}{2}%
\sum_{jk}k_{jk}x_{j}^{0}x_{k}^{0} \label{wqs}%
\end{equation}
where%
\begin{equation}
\text{\ }x_{m}^{0}=\sum_{p}(k^{-1})_{mp}g_{p},m=1,2,\cdots3\mathcal{N}
\label{stad}%
\end{equation}
is the static displacement for the $m^{\text{th}}$ degree of freedom. The
constant force $-\frac{\partial U}{\partial X_{p}}$ exerted by electron on the
$p^{\text{th}}$ vibrational degree of freedom produces a static displacement
$x_{m}^{0}$ for the $m^{\text{th}}$ degree of freedom. The deformation caused
by the static external force of e-ph interaction is balanced by the elastic
force.\ A similar result was obtained for continuum model\cite{Emin76}. The
last term in Eq.(\ref{wqs}) is the polarization energy, a combined
contribution from elastic energy and e-ph interaction. The static displacement
caused by e-ph interaction is meaningful only when the static displacement is
comparable or larger than the thermal vibrational amplitude $\sqrt{\frac
{k_{B}T}{M\omega^{2}}}$ and the zero point vibrational amplitude $\sqrt
{\frac{\hbar\omega}{M\omega^{2}}}$. For a localized state, one needs to make
following substitution%
\begin{equation}
g_{p}\rightarrow d_{p}=\left\{
\begin{array}
[c]{c}%
d_{p_{A_{2}}}\text{ \ \ if \ }p\in D_{A_{2}}\\
0\text{ \ \ if \ }p\notin D_{A_{2}}%
\end{array}
\right.  \label{sub}%
\end{equation}
in corresponding formulae. Because the wave function of a localized state is
concentrated, the e-ph coupling for localized states is treated as a static
displacement. $x_{m}^{0}\thicksim\frac{Z^{\ast}e^{2}}{4\pi\epsilon_{0}\xi
^{2}k},$ $k$ is the typical value of spring constant of a bond. The e-ph
interaction for extended states, which cause scattering among extended states
or from extended state to localized state, is treated as a perturbation.

\subsubsection{Electron transfer}

If there are only two localized states in the \textquotedblleft one electron +
many nuclei\textquotedblright\ system, Eq.(\ref{s11}) is simplified as%
\begin{equation}
i\hbar\frac{\partial}{\partial t}\left(
\begin{array}
[c]{c}%
a_{1}\\
a_{2}%
\end{array}
\right)  =H_{tot}\left(
\begin{array}
[c]{c}%
a_{1}\\
a_{2}%
\end{array}
\right)  ,\text{ }H_{tot}=h_{v}+\left(
\begin{array}
[c]{cc}%
E_{1}^{0}-\sum_{p\in D_{1}}d_{p}x_{p} & J_{12}\\
J_{21} & E_{2}^{0}-\sum_{p\in D_{2}}d_{p}x_{p}%
\end{array}
\right)  \label{ts}%
\end{equation}
Eq.(\ref{ts}) is more general than the usual spin-boson model of electron
transfer. The reason that H$_{tot}$ is not Hermitian is discussed in the
paragraph adjoining Eq.(\ref{tra1}). H$_{tot}$ in (\ref{ts}) can be separated
in a different way:%
\begin{equation}
H_{tot}=H_{EL}+H_{RC}+H_{I}+H_{B} \label{tot}%
\end{equation}
where%
\begin{equation}
H_{EL}=\left(
\begin{array}
[c]{cc}%
E_{1}^{0} & J_{12}\\
J_{21} & E_{2}^{0}%
\end{array}
\right)  \label{etra}%
\end{equation}
is a bare electronic two-state system, and%
\[
H_{RC}=\sum_{j\in D_{1}}[-\frac{\hbar^{2}}{2M_{j}}\nabla_{j}^{2}+\frac{1}%
{2}\sum_{k\in D_{1}}k_{jk}x_{j}x_{k}]+\sum_{j\in D_{2}}[-\frac{\hbar^{2}%
}{2M_{j}}\nabla_{j}^{2}+\frac{1}{2}\sum_{k\in D_{2}}k_{jk}x_{j}x_{k}]
\]%
\begin{equation}
-\sum_{p\in D_{1}}d_{p}x_{p}\left(
\begin{array}
[c]{cc}%
1 & 0\\
0 & 0
\end{array}
\right)  -\sum_{p\in D_{2}}d_{p}x_{p}\left(
\begin{array}
[c]{cc}%
0 & 0\\
0 & 1
\end{array}
\right)  \label{reac}%
\end{equation}
describes the coupling of the two electronic states with the vibrational
degrees of freedom (reaction coordinates) of the two localized states. The
degrees of freedom which belong to $D_{1}$ or $D_{2}$ are reaction
coordinates, removing these degrees of freedom in $h_{v}$, one obtains the
bath Hamiltonian%
\begin{equation}
H_{B}=\sum_{j}{}^{\prime}-\frac{\hbar^{2}}{2M_{j}}\nabla_{j}^{2}+\frac{1}%
{2}\sum_{jk}{}^{\prime}k_{jk}x_{j}x_{k} \label{bath}%
\end{equation}
The primes on the two summation signs indicate that the degrees of freedom of
the atoms inside $D_{1}$ or $D_{2}$ are excluded.%
\begin{equation}
H_{I}=\frac{1}{2}\sum_{k\in D_{1}}\sum_{j\notin D_{1}}k_{jk}x_{j}x_{k}%
+\frac{1}{2}\sum_{k\in D_{2}}\sum_{j\notin D_{2}}k_{jk}x_{j}x_{k}
\label{BRCcou}%
\end{equation}
is the coupling of reaction coordinates with the degrees of freedom of the
thermal bath. Eq.(\ref{tot}) is the starting point of electron transfer
theory\cite{ank}. In this representation, the hopping of an electron between
two localized states appears as that a particle with mass about atomic mass
moves along reaction path between two wells, $H_{B}$ is the environment in
which the particle moves. $H_{I}$ describes the diffusion of the particle and
the friction suffered by the particle.

\subsubsection{Capture and escape of a particle}

If there is one localized state and one extended state, Eqs.(\ref{s11}) and
(\ref{s22}) are simplified to%
\begin{equation}
i\hbar\frac{\partial}{\partial t}\left(
\begin{array}
[c]{c}%
b_{B_{1}}\\
a_{A_{2}}%
\end{array}
\right)  =H_{ce}\left(
\begin{array}
[c]{c}%
b_{B_{1}}\\
a_{A_{2}}%
\end{array}
\right)  ,\text{ }H_{ce}=h_{v}+\left(
\begin{array}
[c]{cc}%
E_{B_{1}} & J_{B_{1}A_{2}}^{\prime}\\
K_{A_{2}B_{1}}^{\prime} & E_{A_{2}}^{0}-\sum_{p\in D_{A_{2}}}d_{p}x_{p}%
\end{array}
\right)  \label{ce}%
\end{equation}
The reason why H$_{ce}$ is not Hermitian is discussed after Eq.(\ref{tra2}).
The capture-escape Hamiltonian H$_{ce}$ can be resolved in a traditional way:%
\begin{equation}
H_{ce}=H_{bf}+H_{RC}^{\prime}+H_{I}^{\prime}+H_{B}^{\prime} \label{ceo}%
\end{equation}
where%
\begin{equation}
H_{bf}=\left(
\begin{array}
[c]{cc}%
E_{B_{1}} & J_{B_{1}A_{2}}^{\prime}\\
K_{A_{2}B_{1}}^{\prime} & E_{A_{2}}^{0}%
\end{array}
\right)  \label{et}%
\end{equation}
is a bare electronic two-state system: one is the localized state, another is
the extended state.%
\begin{equation}
H_{RC}^{\prime}=\sum_{j\in D_{A_{2}}}[-\frac{\hbar^{2}}{2M_{j}}\nabla_{j}%
^{2}+\frac{1}{2}\sum_{k\in D_{A_{2}}}k_{jk}x_{j}x_{k}]-\sum_{p\in D_{A_{2}}%
}d_{p}x_{p}\left(
\begin{array}
[c]{cc}%
0 & 0\\
0 & 1
\end{array}
\right)  \label{espa}%
\end{equation}
describes the coupling of the two electronic states with the vibrational
degrees of freedom (reaction coordinates) of the localized state. The degrees
of freedom which belong to $D_{2}$ are reaction coordinates, removing these
degrees of freedom in $h_{v}$, one obtains the bath Hamiltonian%
\begin{equation}
H_{B}^{\prime}=\sum_{j}{}^{\prime}-\frac{\hbar^{2}}{2M_{j}}\nabla_{j}%
^{2}+\frac{1}{2}\sum_{jk}{}^{\prime}k_{jk}x_{j}x_{k} \label{ene}%
\end{equation}
The primes on the two summation signs indicate that the degrees of freedom of
the atoms inside $D_{2}$ are excluded.%
\begin{equation}
H_{I}^{\prime}=\frac{1}{2}\sum_{k\in D_{A_{2}}}\sum_{j\notin D_{A_{2}}}%
k_{jk}x_{j}x_{k} \label{ens}%
\end{equation}
is the coupling of reaction coordinates with the degrees of freedom of the
thermal bath.

In this representation, the transition of an electron from a localized state
to an extended state appears as that a particle with mass about atomic mass
diffuses in a viscous liquid like a Brownian particle and escapes the trap of
a well. Taking a semi-classical approximation, one obtains Kramers' problem of
particle escape across a barrier\cite{kram}. The transition from an extended
state to a localized state appears as that a well captures a particle moving
in a viscous liquid.

\subsection{Normal Coordinates}

As usual it is convenient to change the displacements $\{x\}$ of the atoms to
normal coordinates\cite{dra,tafn} $\{\Theta\}$,%

\begin{equation}
x_{k}=\sum_{\alpha}\Delta_{k\alpha}\Theta_{\alpha}\text{, \ \ \ }(\Delta
^{T}k\Delta)_{\beta\alpha}=\delta_{\alpha\beta}M_{\alpha}\omega_{\alpha}%
^{2},\text{ \ \ }\alpha=1,2,\cdots,3\mathcal{N} \label{nor1}%
\end{equation}
where $\Delta_{k\alpha}$ is the minor of the determinant $\left\vert
k_{ik}-\omega^{2}M_{i}\delta_{ik}\right\vert =0$, $\Delta^{T}$ is the
transpose matrix of the matrix $(\Delta_{k\alpha})$. The two coupling
constants in Eq.(\ref{ept}) and Eq.(\ref{ep0}) which involve e-ph interaction
are expressed as%
\begin{equation}
K_{A_{2}B_{1}}^{\prime}=\sum_{\alpha}\Theta_{\alpha}K_{A_{2}B_{1}}%
^{\prime\alpha},\text{ \ \ }K_{A_{2}B_{1}}^{\prime\alpha}=\sum_{j}%
\Delta_{j\alpha}\int d^{3}r\phi_{A_{2}}^{\ast}\frac{\partial U}{\partial
X_{j}}\xi_{B_{1}} \label{kp}%
\end{equation}
and
\begin{equation}
K_{B_{2}B_{1}}=\sum_{\alpha}\Theta_{\alpha}K_{B_{2}B_{1}}^{\alpha},\text{
\ }K_{B_{2}B_{1}}^{\alpha}=\sum_{j}\Delta_{j\alpha}\int d^{3}r\xi_{B_{2}%
}^{\ast}\frac{\partial U}{\partial X_{j}}\xi_{B_{1}} \label{k}%
\end{equation}
where $K_{A_{2}B_{1}}^{\prime\alpha}$ and $K_{A_{2}B_{1}}^{\alpha}$have the
dimension of force. Eq.(\ref{s11}) and Eq.(\ref{s22}) become:%
\begin{equation}
(i\hbar\frac{\partial}{\partial t}-h_{A_{2}})a_{A_{2}}(\cdots\Theta_{\alpha
}\cdots;t)=\sum_{A_{1}}J_{A_{2}A_{1}}a_{A_{1}}+\sum_{B_{1}}K_{A_{2}B_{1}%
}^{\prime}b_{B_{1}} \label{1s}%
\end{equation}
and
\begin{equation}
(i\hbar\frac{\partial}{\partial t}-h_{B_{2}})b_{B_{2}}(\cdots\Theta_{\alpha
}\cdots;t)=\sum_{A_{1}}J^{\prime}{}_{B_{2}A_{1}}a_{A_{1}}+\sum_{B_{1}}%
K_{B_{2}B_{1}}b_{B_{1}} \label{2s}%
\end{equation}
where%
\begin{equation}
h_{A_{1}}=E_{A_{1}}+h_{v},\text{ \ \ \ \ }h_{B_{2}}=E_{B_{2}}+h_{v} \label{h1}%
\end{equation}
$h_{A_{1}}$ describes the polarization on the amorphous network caused by an
electron in localized state $\phi_{A_{1}}$ through e-ph coupling. $a_{A_{2}%
}(\cdots\Theta_{\alpha}\cdots;t)$ is the probability amplitude at moment $t$
that the electron is in localized state $A_{2}$ while the vibrational state of
the atoms is given by normal coordinates $\{\Theta_{\alpha},\alpha
=1,2,\cdots,3\mathcal{N}\}$. $b_{B_{2}}(\cdots\Theta_{\alpha}\cdots;t)$ is the
probability amplitude at moment $t$ that the electron is in extended state
$B_{2}$ while the vibrational state of the atoms is given by normal
coordinates $\{\Theta_{\alpha},\alpha=1,2,\cdots,3\mathcal{N}\}$.

Due to the coupling of localized state $A_{1}$ with the vibrations of atoms,
by a similar reasoning as Eq.(\ref{stad}), the origin of each normal
coordinate is shifted\cite{lee,LLP}%
\begin{equation}
\Theta_{\alpha}\rightarrow\Theta_{\alpha}-\Theta_{\alpha}^{A_{1}},\text{
\ \ \ \ }\Theta_{\alpha}^{A_{1}}=\left(  M_{\alpha}\omega_{\alpha}^{2}\right)
^{-1}\sum_{p_{A_{1}}\mathbf{\in}D_{A_{1}}}d_{p_{A_{1}}}\Delta_{p_{A_{1}}%
\alpha}, \label{or}%
\end{equation}
\ where $\Theta_{\alpha}^{A_{1}}\thicksim\frac{N_{A_{1}}}{\mathcal{N}}%
\frac{Z^{\ast}e^{2}}{4\pi\epsilon_{0}\xi_{A_{1}}^{2}\left(  M_{\alpha}%
\omega_{\alpha}^{2}\right)  }$ is the static displacement in normal coordinate
of the $\alpha^{\text{th}}$ mode caused by the coupling with localized state
$A_{1}$, where $N_{A_{1}}$ is the number of atoms in region $D_{A_{1}}$.
Eq.(\ref{or}) leads to a modification to the phonon wave function (cf.
Eq.(\ref{ew})) and a change in total energy (cf. Eq.(\ref{ev})). Using
$(k^{-1})_{jk}=(k^{-1})_{kj}$ and the inverse relations of Eq.(\ref{nor1}),
one finds that the shift $\Theta_{\alpha}^{A_{1}}$ of origin of the
$\alpha^{\text{th}}$ normal coordinate is related to the static displacements
by%
\begin{equation}
\Theta_{\alpha}^{A_{1}}=\sum_{k}(\Delta^{-1})_{\alpha k}x_{k}^{0},\text{
\ }x_{k}^{0}\in D_{A_{1}} \label{shif}%
\end{equation}

The eigenfunctions of $h_{A_{1}}$ are%
\begin{equation}
\Psi_{A_{1}}^{\{N_{\alpha}\}}=%
{\displaystyle\prod\limits_{\alpha=1}^{3\mathcal{N}}}
\Phi_{N_{\alpha}}(\theta_{\alpha}-\theta_{\alpha}^{A_{1}}),\text{ \ }\Phi
_{N}(z)=(2^{N}N!\pi^{1/2})^{-1/2}e^{-z^{2}/2}H_{N}(z), \label{ew}%
\end{equation}
where $H_{N}(z)$ is the $N^{\text{th}}$ Hermite polynomial, $\theta_{\alpha
}=(\frac{M_{\alpha}\omega_{\alpha}}{\hbar})^{1/2}\Theta_{\alpha}$ is the
dimensionless normal coordinate and $\theta_{\alpha}^{A_{1}}=(\frac{M_{\alpha
}\omega_{\alpha}}{\hbar})^{1/2}\Theta_{\alpha}^{A_{1}}$. The corresponding
eigenvalues are%
\begin{equation}
\mathcal{E}_{A_{1}}^{\{N_{\alpha}\}}=E_{A_{1}}^{0}+\sum_{\alpha}(N_{\alpha
}+\frac{1}{2})\hbar\omega_{\alpha}+\mathcal{E}_{A_{1}}^{b},\text{
\ }\mathcal{E}_{A_{1}}^{b}=-\frac{1}{2}\sum_{\alpha}M_{\alpha}\omega_{\alpha
}^{2}(\Theta_{\alpha}^{A_{1}})^{2} \label{ev}%
\end{equation}
$\mathcal{E}_{A_{1}}^{b}\thicksim k^{-1}[\frac{Z^{\ast}e^{2}}{4\pi\epsilon
_{0}\xi_{A_{1}}^{2}}]^{2}$. In a small polaron, the electron is mainly
localized on \textit{one} site. The localized electron deforms the crystalline
lattice. In an amorphous semiconductor, the electron in state $A_{1}$
polarizes the network, the energy of state $A_{1}$ is shifted downward by
$\mathcal{E}_{A_{1}}^{b}$. Eq.(\ref{ev}) is an extension of Holstein's small
polaron theory. Observe that $\mathcal{E}_{A_{1}}^{b}$ does not depend on the
number of phonons in each mode and is a static property of the amorphous
solid. $\mathcal{E}_{A_{1}}^{b}$ was derived in usual Born-Oppenheimer
approach and is called \textquotedblleft dynamic potential
energy\textquotedblright\cite{chr}. With the help of Eq.(\ref{stad}) and the
inverse relation of Eq.(\ref{nor1}), one can show that the two expressions for
the polarization energy are the same:%
\begin{equation}
\frac{1}{2}\sum_{jk}k_{jk}x_{j}^{0}x_{k}^{0}=\frac{1}{2}\sum_{pq}d_{p}%
d_{q}(k^{-1})_{qp}=|\mathcal{E}_{A_{1}}^{b}| \label{pole}%
\end{equation}
The continuum form of Eq.(\ref{pole}) is Eq.(6c) of ref.\onlinecite{Emin76}.
The eigenvalues and eigenvectors of $h_{B_{1}}$ are%
\begin{equation}
\mathcal{E}_{B_{1}}^{\{N_{\alpha}\}}=E_{B_{1}}+\sum_{\alpha}(N_{\alpha}%
+\frac{1}{2})\hbar\omega_{\alpha},\text{ \ \ \ \ }\Xi_{A_{1}}^{\{N_{\alpha}%
\}}=%
{\displaystyle\prod\limits_{\alpha=1}^{3\mathcal{N}}}
\Phi_{N_{\alpha}}(\theta_{\alpha}) \label{bv}%
\end{equation}

We change probability amplitude $a_{A_{1}}(\cdots\Theta_{\alpha}\cdots;t)$
from representation of normal coordinates to representation of occupation
number, i.e. expand with the eigenfunctions of $h_{A_{1}}$
\begin{equation}
a_{A_{1}}=\sum_{\cdots N_{\alpha}^{\prime}\cdots}C^{A_{1}}(\cdots N_{\alpha
}^{\prime}\cdots;t)\Psi_{A_{1}}^{\{N_{\alpha}^{\prime}\}}e^{-it\mathcal{E}%
_{A_{1}}^{\{N_{\alpha}^{\prime}\}}/\hbar} \label{az}%
\end{equation}
$C^{A_{1}}(\cdots N_{\alpha}^{\prime}\cdots;t)$ is the probability amplitude
at moment $t$ that the electron is in localized state $A_{1}$ while the
vibrational state of the nuclei is characterized by occupation number
$\{N_{\alpha}^{\prime},\alpha=1,2,\cdots,3\mathcal{N}\}$ in each mode.
Similarly we expand the probability amplitude $b_{B_{1}}(\cdots\Theta_{\alpha
}\cdots;t)$ with eigenfunctions of $h_{B_{1}}$%
\begin{equation}
b_{B_{1}}=\sum_{\cdots N_{\alpha}^{\prime}\cdots}F^{B_{1}}(\cdots N_{\alpha
}^{\prime}\cdots;t)\Xi_{B_{1}}^{\{N_{\alpha}^{\prime}\}}e^{-it\mathcal{E}%
_{B_{1}}^{\{N_{\alpha}^{\prime}\}}/\hbar} \label{bz}%
\end{equation}
$F^{B_{1}}(\cdots N_{\alpha}^{\prime}\cdots;t)$ is the probability amplitude
at moment $t$ that the electron is in extended state $B_{1}$ while the
vibrational state of the nuclei is characterized by occupation number
$\{N_{\alpha}^{\prime},\alpha=1,2,\cdots,3\mathcal{N}\}$ in each mode.

Substitute Eq.(\ref{az}) and Eq.(\ref{bz}) into Eq.(\ref{1s}) and applying
$\int%
{\displaystyle\prod\limits_{\alpha}}
d\theta_{\alpha}\Psi_{A_{2}}^{\{N_{\alpha}\}}$ to both sides we obtain%
\begin{equation}
i\hbar\frac{\partial C^{A_{2}}}{\partial t}=\sum_{A_{1}\cdots N_{\alpha
}^{\prime}\cdots}\langle A_{2}\cdots N_{\alpha}\cdots|V_{LL}^{tr}|A_{1}\cdots
N_{\alpha}^{\prime}\cdots\rangle C^{A_{1}}(\cdots N_{\alpha}^{\prime}%
\cdots;t)e^{it(\mathcal{E}_{A_{2}}^{\{N_{\alpha}\}}-\mathcal{E}_{A_{1}%
}^{\{N_{\alpha}^{\prime}\}})/\hbar} \label{cq}%
\end{equation}%
\[
+\sum_{B_{1}\cdots N_{\alpha}^{\prime}\cdots}\langle A_{2}\cdots N_{\alpha
}\cdots|V_{EL}^{e-ph}|B_{1}\cdots N_{\alpha}^{\prime}\cdots\rangle F^{B_{1}%
}(\cdots N_{\alpha}^{\prime}\cdots;t)e^{it(\mathcal{E}_{A_{2}}^{\{N_{\alpha
}\}}-\mathcal{E}_{B_{1}}^{\{N_{\alpha}^{\prime}\}})/\hbar}%
\]
where%
\begin{equation}
\langle A_{2}\cdots N_{\alpha}\cdots|V_{LL}^{tr}|A_{1}\cdots N_{\alpha
}^{\prime}\cdots\rangle=J_{A_{2}A_{1}}\int%
{\displaystyle\prod\limits_{\alpha}}
d\theta_{\alpha}\Psi_{A_{2}}^{\{N_{\alpha}\}}\Psi_{A_{1}}^{\{N_{\alpha
}^{\prime}\}} \label{LLT}%
\end{equation}
describes the transition from localized state $A_{1}$ with phonon distribution
$\{\cdots N_{\alpha}^{\prime}\cdots\}$ to localized state $A_{2}$ with phonon
distribution $\{\cdots N_{\alpha}\cdots\}$ caused by transfer integral
$J_{A_{2}A_{1}}$ defined in Eq.(\ref{tra1}).%
\begin{equation}
\langle A_{2}\cdots N_{\alpha}\cdots|V_{EL}^{e-ph}|B_{1}\cdots N_{\alpha
}^{\prime}\cdots\rangle=\int%
{\displaystyle\prod\limits_{\alpha}}
d\theta_{\alpha}\Psi_{A_{2}}^{\{N_{\alpha}\}}(\sum_{\alpha}\Theta_{\alpha
}K_{A_{2}B_{1}}^{\prime\alpha})\Xi_{B_{1}}^{\{N_{\alpha}^{\prime}\}}
\label{ELT}%
\end{equation}
is the transition from an extended state to a localized state induced by
electron-phonon interaction.

Similarly from Eq.(\ref{2s}) we have%
\begin{equation}
i\hbar\frac{\partial F_{\{N_{\alpha}\}}^{B_{2}}}{\partial t}=\sum_{A_{1}\cdots
N_{\alpha}^{\prime}\cdots}\langle B_{2}\cdots N_{\alpha}\cdots|V_{LE}%
^{tr}|A_{1}\cdots N_{\alpha}^{\prime}\cdots\rangle C_{\{N_{\alpha}^{\prime}%
\}}^{A_{1}}e^{it(\mathcal{E}_{B_{2}}^{\{N_{\alpha}\}}-\mathcal{E}_{A_{1}%
}^{\{N_{\alpha}^{\prime}\}})/\hbar} \label{fq}%
\end{equation}%
\[
+\sum_{B_{1}\cdots N_{\alpha}^{\prime}\cdots}\langle B_{2}\cdots N_{\alpha
}\cdots|V_{EE}^{e-ph}|B_{1}\cdots N_{\alpha}^{\prime}\cdots\rangle
F_{\{N_{\alpha}^{\prime}\}}^{B_{1}}e^{it(\mathcal{E}_{B_{2}}^{\{N_{\alpha}%
\}}-\mathcal{E}_{B_{1}}^{\{N_{\alpha}^{\prime}\}})/\hbar}%
\]
where%
\begin{equation}
\langle B_{2}\cdots N_{\alpha}\cdots|V_{LE}^{tr}|A_{1}\cdots N_{\alpha
}^{\prime}\cdots\rangle=J^{\prime}{}_{B_{2}A_{1}}\int%
{\displaystyle\prod\limits_{\alpha}}
d\theta_{\alpha}\Xi_{B_{2}}^{\{N_{\alpha}\}}\Psi_{A_{1}}^{\{N_{\alpha}%
^{\prime}\}} \label{LET}%
\end{equation}
describes the transition from localized state $|A_{1}\cdots N_{\alpha}%
^{\prime}\cdots\rangle$ to extended state $|B_{2}\cdots N_{\alpha}%
\cdots\rangle$ caused by transfer integral $J^{\prime}{}_{B_{2}A_{1}}$, the
dependence on $\{x_{j}\}$ in $J^{\prime}$ is neglected.
\begin{equation}
\langle B_{2}\cdots N_{\alpha}\cdots|V_{EE}^{e-ph}|B_{1}\cdots N_{\alpha
}^{\prime}\cdots\rangle=\int%
{\displaystyle\prod\limits_{\alpha}}
d\theta_{\alpha}\Xi_{B_{2}}^{\{N_{\alpha}\}}(\sum_{\alpha}\Theta_{\alpha
}K_{B_{2}B_{1}}^{\alpha})\Xi_{B_{1}}^{\{N_{\alpha}^{\prime}\}} \label{ee1}%
\end{equation}
is the matrix element of the transition between two extended states caused by
electron-phonon interaction. It is similar to the usual expression in a metal.
Eq.(\ref{cq}) and Eq.(\ref{fq}) are the evolution equations in
second-quantized form. The phonon state on the left hand side (LHS) can be
different from that in the right hand side. In general, the occupation number
in each mode changes when the electron changes its state.

\section{Transition between two localized states}

If we only consider localized states, Eq.(\ref{cq}) is simplified to%
\begin{equation}
i\hbar\frac{\partial C^{A_{3}}}{\partial t}=\sum_{A_{1}\cdots N_{\alpha
}^{\prime}\cdots}\langle A_{3}\cdots N_{\alpha}\cdots|V_{LL}^{tr}|A_{1}\cdots
N_{\alpha}^{\prime}\cdots\rangle C^{A_{1}}(\cdots N_{\alpha}^{\prime}%
\cdots)e^{it(\mathcal{E}_{A_{3}}^{\{N_{\alpha}\}}-\mathcal{E}_{A_{1}%
}^{\{N_{\alpha}^{\prime}\}})/\hbar} \label{c3}%
\end{equation}
Eq.(\ref{1s}) is reduced to%
\begin{equation}
(i\hbar\frac{\partial}{\partial t}-h_{A_{3}})a_{A_{3}}(\cdots\Theta_{\alpha
}\cdots)=\sum_{A_{1}}J_{A_{3}A_{1}}a_{A_{1}} \label{LLn}%
\end{equation}
Eq.(\ref{c3}) corresponds to the small polaron problem\cite{Hol1,Hol2,Emin75}
and Eq.(\ref{LLn}) corresponds to the electron transfer among ions in polar
solvent\cite{msta,met,et}. The transfer integral $J_{A_{3}A_{1}}$ decays
exponentially with the distance between the states $A_{1}$ and $A_{3}$.
According to Eq.(\ref{or}), the shift of origin $\Theta_{\alpha}^{A_{3}}$ for
the $\alpha^{\text{th}}$ mode is order of $\frac{N_{A_{3}}}{\mathcal{N}},$
$\mathcal{N}$ is the number of atoms in sample, $N_{A_{3}}$ is the number of
atoms in $D_{A_{3}}$. If $A_{3}$ and $A_{1}$ are not close to the mobility
edge, $(\theta_{\alpha}^{A_{3}}-\theta_{\alpha}^{A_{1}})$ is infinitesimal. To
compute the integral in Eq.(\ref{LLT}), we expand $\Phi_{N_{\alpha}^{\prime}%
}(\theta_{\alpha}-\theta_{\alpha}^{A_{1}})$ around $(\theta_{\alpha}%
-\theta_{\alpha}^{A_{3}})$ to second order in $(\theta_{\alpha}^{A_{3}}%
-\theta_{\alpha}^{A_{1}})$. To the second order of the small parameter
$(\theta_{\alpha}^{A_{3}}-\theta_{\alpha}^{A_{1}})$, the integral for the
$\alpha^{\text{th}}$ mode in Eq.(\ref{LLT}) can be effected.

\subsection{$J_{_{A_{3}A_{1}}}$ as perturbation}

If the transfer integral $J_{_{A_{3}A_{1}}}$ is not small, the semi-classical
Fokker-Planck equation is an useful approximation\cite{et} for solving
Eq.(\ref{LLn}). However in amorphous solids, the transfer integral
Eq.(\ref{tra1}) between two localized states can be small. Perturbation theory
can be used to compute the transition \ probability\cite{Hol2} $W$ from state
$\Psi_{\{N_{\alpha}^{\prime}\}}^{A_{1}}$ to state $\Psi_{\{N_{\alpha}%
\}}^{A_{3}}$. If in the initial instant ($t=0$) the \textquotedblleft one
electron + many nuclei\textquotedblright\ system is at state $|A_{1}\cdots
N_{\alpha}^{\prime}\cdots\rangle$, then only $C^{A_{1}}(\cdots N_{\alpha
}^{\prime}\cdots;t=0)=1$ and other coefficients are zero. At later moment $t$,
probability amplitude $C^{A_{3}}(A_{3}\cdots N_{\alpha}\cdots,t)$ is simply
determined by Eq.(\ref{c3}). The transition probability per unit time is%
\begin{equation}
W(A_{1}\cdots N_{\alpha}^{\prime}\cdots\rightarrow A_{3}\cdots N_{\alpha
}\cdots)=\frac{1}{\hbar^{2}}|\langle A_{3}\cdots N_{\alpha}\cdots|V_{LL}%
^{tr}|A_{1}\cdots N_{\alpha}^{\prime}\cdots\rangle|^{2} \label{pro}%
\end{equation}%
\[
\frac{\partial}{\partial t}\int_{0}^{t}dt^{\prime}\int_{-t^{\prime}%
}^{t^{\prime}}dt^{\prime\prime}\exp\{\frac{it^{\prime\prime}}{\hbar
}(\mathcal{E}_{\{N_{\alpha}\mathcal{\}}}^{A_{3}}-\mathcal{E}_{\{N_{\alpha
}^{\prime}\mathcal{\}}}^{A_{1}})\}
\]
Because there exist low frequency acoustic modes in any solid, limit
$t\rightarrow\infty$ must be taken at a later stage\cite{Hol2}. For two
localized states, transfer integral (\ref{tra1}) does not satisfy
$J_{A_{2}A_{1}}=(J_{A_{1}A_{2}})^{\ast}$. One must carefully distinguish the
transition $|A_{1}\cdots N_{\alpha}^{\prime}\cdots\rangle\rightarrow
|A_{3}\cdots N_{\alpha}\cdots\rangle$ which is caused by the attraction on the
electron by the atoms in $D_{A_{3}}$ and its inverse transition $|A_{3}\cdots
N_{\alpha}\cdots\rangle\rightarrow|A_{1}\cdots N_{\alpha}^{\prime}%
\cdots\rangle$ which is caused by the attraction on the electron by the atoms
in $D_{A_{1}}$. In the usual situation, the perturbation which leads to the
forward transition is the same as the perturbation which leads to the backward
transition. The two transition probabilities equal. If the spatial regions of
two localized states have many overlaps, the difference in the transition
probabilities between the forward and the backward direction becomes small.

Substitute Eq.(\ref{ev}) and Eq.(\ref{LLT}) into Eq.(\ref{pro}), notice that
the product of two conflicting Kronecker delta symbols is zero, the transition
probability becomes%
\begin{equation}
W(A_{1}\cdots N_{\alpha}^{\prime}\cdots\rightarrow A_{3}\cdots N_{\alpha
}\cdots)=\frac{J_{A_{3}A_{1}}^{2}}{\hbar^{2}}\int_{-t}^{t}dt^{\prime}%
\exp\{\frac{it^{\prime}}{\hbar}[(E_{A_{1}}^{0}+\mathcal{E}_{b}^{A_{1}%
})-(E_{A_{3}}^{0}+\mathcal{E}_{b}^{A_{3}})]\} \label{sep}%
\end{equation}%
\[%
{\displaystyle\prod\limits_{\alpha}}
\{\delta_{N_{\alpha}^{\prime}N_{\alpha}}e^{it^{\prime}\omega_{\alpha
}(N_{\alpha}^{\prime}-N_{\alpha})}[1-(N_{\alpha}^{\prime}+\frac{1}{2}%
)(\theta_{\alpha}^{A_{3}}-\theta_{\alpha}^{A_{1}})^{2}]
\]%
\[
+(\theta_{\alpha}^{A_{3}}-\theta_{\alpha}^{A_{1}})^{2}(\frac{N_{\alpha
}^{\prime}}{2})\delta_{N_{\alpha}^{\prime}-1,N_{\alpha}}e^{it^{\prime}%
\omega_{\alpha}(N_{\alpha}^{\prime}-N_{\alpha})}+(\theta_{\alpha}^{A_{3}%
}-\theta_{\alpha}^{A_{1}})^{2}(\frac{N_{\alpha}^{\prime}+1}{2})\delta
_{N_{\alpha}^{\prime}+1,N_{\alpha}}e^{it^{\prime}\omega_{\alpha}(N_{\alpha
}^{\prime}-N_{\alpha})}\}
\]
where terms order of $(\theta_{\alpha}^{A_{3}}-\theta_{\alpha}^{A_{1}})^{4}$
and higher are neglected.

Next we sum over all possible final phonon states $\{\cdots N_{\alpha}%
\cdots\}$ and take a thermal average over initial phonon states $\{\cdots
N_{\alpha}^{\prime}\cdots\}$, the transition probability from $A_{1}$ to
$A_{3}$ is
\[
W(A_{1}\rightarrow A_{3})=Z^{-1}\sum_{\cdots N_{\alpha}^{\prime}\cdots
}W_{\cdots N_{\alpha}^{\prime}\cdots}(A_{1}\rightarrow A_{3})\exp[-\beta
\sum_{\alpha}(N_{\alpha}^{\prime}+\frac{1}{2})\hbar\omega_{\alpha}]
\]%
\[
=\frac{J_{A_{3}A_{1}}^{2}}{\hbar^{2}}\int_{-t}^{t}dt^{\prime}\exp
\{\frac{it^{\prime}}{\hbar}[(E_{A_{3}}^{0}+\mathcal{E}_{b}^{A_{3}})-(E_{A_{1}%
}^{0}+\mathcal{E}_{b}^{A_{1}})]\}
\]%
\begin{equation}
\exp\{-\frac{1}{2}\sum_{\alpha}(\theta_{\alpha}^{A_{3}}-\theta_{\alpha}%
^{A_{1}})^{2}[\coth\frac{^{\beta\hbar\omega_{\alpha}}}{2}(1-\cos t^{\prime
}\omega_{\alpha})-i\sin t^{\prime}\omega_{\alpha}]\} \label{t}%
\end{equation}
where $Z=\sum_{\cdots N_{\alpha}^{\prime}\cdots}\exp[-\beta\sum_{\alpha
}(N_{\alpha}^{\prime}+\frac{1}{2})\hbar\omega_{\alpha}]$ is the partition
function of phonon. Here we implicitly assumed that the vibrational state of
atoms stays in equilibrium, and is not affected by the motion of other electrons.

Because $\sum_{\alpha}(\theta_{\alpha}^{A_{3}}-\theta_{\alpha}^{A_{1}}%
)^{2}\thicksim\mathcal{N}$ is large and the integrand is analytic about
$t^{\prime}$, the integral in Eq.(\ref{t}) may be estimated by the method of
steepest descent. Let the time derivative of the exponent of last exponential
of Eq.(\ref{t}) equal zero, one has $\coth\frac{^{\beta\hbar\omega_{\alpha}}%
}{2}=i\cot t^{\prime}\omega_{\alpha}$. The saddle point is $t_{s}^{\prime
}=\frac{i\beta\hbar}{2}$. To effect the time integral in Eq.(\ref{t}), change
variable\cite{Hol2} from $t^{\prime}$ to $\tau:$ $t^{\prime}=\frac{i\beta
\hbar}{2}+\tau$. The path of integral is also changed $\int_{-t}^{t}%
dt^{\prime}\Longrightarrow\int_{-t-\frac{i\beta\hbar}{2}}^{-t}d\tau+\int
_{-t}^{t}d\tau+\int_{t}^{t-\frac{i\beta\hbar}{2}}d\tau$, Eq.(\ref{t})
becomes:
\begin{equation}
W(A_{1}\rightarrow A_{3})=\frac{J_{A_{3}A_{1}}^{2}}{\hbar^{2}}[g_{1}%
(t)+g_{2I}(t)+g_{2II}(t)]\exp\{\frac{-\beta}{2}[(E_{A_{3}}^{0}+\mathcal{E}%
_{b}^{A_{3}})-(E_{A_{1}}^{0}+\mathcal{E}_{b}^{A_{1}})]\} \label{g}%
\end{equation}
where%
\[
g_{1}(t)=\int_{-t}^{t}d\tau\exp\{\frac{i\tau}{\hbar}[(E_{A_{3}}^{0}%
+\mathcal{E}_{b}^{A_{3}})-(E_{A_{1}}^{0}+\mathcal{E}_{b}^{A_{1}})]\}
\]%
\begin{equation}
\exp\{-\frac{1}{2}\sum_{\alpha}(\theta_{\alpha}^{A_{3}}-\theta_{\alpha}%
^{A_{1}})^{2}(\coth\frac{^{\beta\hbar\omega_{\alpha}}}{2}-\text{csch}%
\frac{\beta\hbar\omega_{\alpha}}{2}\cos\tau\omega_{\alpha})\} \label{g1}%
\end{equation}
$g_{2I}(t)$ and $g_{2II}(t)$ are obtained by taking the upper sign and the
lower sign in%
\[
\pm i\exp\{\frac{-it}{\hbar}[(E_{A_{3}}^{0}+\mathcal{E}_{b}^{A_{3}}%
)-(E_{A_{1}}^{0}+\mathcal{E}_{b}^{A_{1}})]\}\int_{0}^{\frac{\beta\hbar}{2}%
}ds\exp\{\frac{s}{\hbar}[(E_{A_{3}}^{0}+\mathcal{E}_{b}^{A_{3}})-(E_{A_{1}%
}^{0}+\mathcal{E}_{b}^{A_{1}})]\}
\]%
\begin{equation}
\exp\{-\frac{1}{2}\sum_{\alpha}(\theta_{\alpha}^{A_{3}}-\theta_{\alpha}%
^{A_{1}})^{2}(\coth\frac{^{\beta\hbar\omega_{\alpha}}}{2}-\text{csch}%
\frac{\beta\hbar\omega_{\alpha}}{2}\cos(t\pm is)\omega_{\alpha})\}\overset
{t>>1/\overline{\omega}}{\rightarrow}t^{-1/2} \label{2I}%
\end{equation}
respectively. Since we are only interested in `long' transition time,
$g_{2I}(t)$ and $g_{2II}(t)$ will be discarded hereafter\cite{Hol2}.

If the initial phonon state is the same as the final phonon state,
Eq.(\ref{sep}) is changed into%
\begin{equation}
W_{T}^{(d)}(A_{1}\rightarrow A_{3})=\frac{J_{A_{3}A_{1}}^{2}}{\hbar^{2}}%
\exp\{\frac{-\beta}{2}[(E_{A_{3}}^{0}+\mathcal{E}_{b}^{A_{3}})-(E_{A_{1}}%
^{0}+\mathcal{E}_{b}^{A_{1}})]\} \label{dia}%
\end{equation}%
\[
\int_{-t}^{t}d\tau\exp\{\frac{i\tau}{\hbar}[(E_{A_{3}}^{0}+\mathcal{E}%
_{b}^{A_{3}})-(E_{A_{1}}^{0}+\mathcal{E}_{b}^{A_{1}})]\}\exp\{-\frac{1}{2}%
\sum_{\alpha}(\theta_{\alpha}^{A_{3}}-\theta_{\alpha}^{A_{1}})^{2}\coth
\frac{^{\beta\hbar\omega_{\alpha}}}{2}\}
\]
As a counterpart of Eq.(\ref{g}), Eq.(\ref{dia}) is only an intermediate step
in effecting the time integral and does not have any physical meaning. In
deriving Eq.(\ref{g}), i.e. the steps of summing over all final phonon states
and thermal average the initial phonon states, both $(N_{1}^{\prime}\cdots
N_{3\mathcal{N}}^{\prime})$ and $(N_{1}\cdots N_{3\mathcal{N}})$ are treated
as independent variables. Subtracting Eq.(\ref{dia}) from Eq.(\ref{g}), the
phonon assisted transition probability from $A_{3}$ to $A_{1}$ is
\begin{equation}
W_{T}(A_{1}\rightarrow A_{3})=\frac{J_{A_{3}A_{1}}^{2}}{\hbar^{2}}\exp
\{\frac{-\beta}{2}[(E_{A_{3}}^{0}+\mathcal{E}_{b}^{A_{3}})-(E_{A_{1}}%
^{0}+\mathcal{E}_{b}^{A_{1}})]\} \label{TT}%
\end{equation}%
\[
\exp\{-\frac{1}{2}\sum_{\alpha}(\theta_{\alpha}^{A_{3}}-\theta_{\alpha}%
^{A_{1}})^{2}\coth\frac{^{\beta\hbar\omega_{\alpha}}}{2}\}
\]%
\[
\int_{-t}^{t}d\tau\exp\{\frac{i\tau}{\hbar}[(E_{A_{3}}^{0}+\mathcal{E}%
_{b}^{A_{3}})-(E_{A_{1}}^{0}+\mathcal{E}_{b}^{A_{1}})]\}[\exp\{\frac{1}{2}%
\sum_{\alpha}(\theta_{\alpha}^{A_{3}}-\theta_{\alpha}^{A_{1}})^{2}%
\text{csch}\frac{\beta\hbar\omega_{\alpha}}{2}\cos\tau\omega_{\alpha})\}-1]
\]

\subsection{High temperature limit}

\begin{figure}[h]
\begin{center}
\resizebox{90mm}{!}{\includegraphics{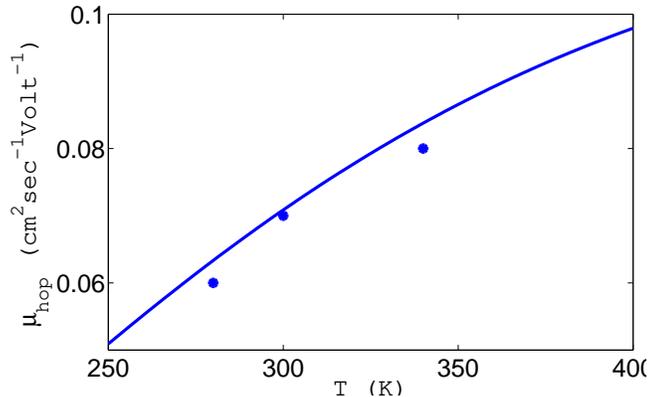}}
\end{center}
\par
.\caption{(Color online) Mobility from LL transition in a-Si (T$>$ 250K):
solid line is computed from Eq.(\ref{Mar}) with parameters: $\lambda
_{LL}=0.2eV,~\Delta G_{LL}=0.05eV,~J_{LL}=0.02eV$. The star symbols are
experimental data\cite{mo,tie,cer,koc}.}%
\label{fig1}%
\end{figure}Even at moderately high temperature $k_{B}T\gtrsim\hbar
\overline{\omega}$, the contribution from diagonal transition, i.e. 1 in last
square bracket can be neglected. The main contribution of the `time' integral
comes from\cite{Hol2} the neighborhood of $\tau=0$. Expand $\cos\tau
\omega_{\alpha}\thickapprox1-\tau^{2}\omega_{\alpha}^{2}/2$ and effect the
Gaussian integral, Eq.(\ref{TT}) is changed into%
\begin{equation}
W_{T}(A_{1}\rightarrow A_{3})=\frac{J_{A_{3}A_{1}}^{2}}{\hbar^{2}}\exp
\{\frac{-\beta}{2}[(E_{A_{3}}^{0}+\mathcal{E}_{b}^{A_{3}})-(E_{A_{1}}%
^{0}+\mathcal{E}_{b}^{A_{1}})]\} \label{go}%
\end{equation}%
\[
\exp\{-\frac{1}{2}\sum_{\alpha}(\theta_{\alpha}^{A_{3}}-\theta_{\alpha}%
^{A_{1}})^{2}\tanh\frac{^{\beta\hbar\omega_{\alpha}}}{4}\}
\]%
\[
(2\pi)^{1/2}[\frac{1}{2}\sum_{\alpha}(\theta_{\alpha}^{A_{3}}-\theta_{\alpha
}^{A_{1}})^{2}\omega_{\alpha}^{2}\text{csch}\frac{\beta\hbar\omega_{\alpha}%
}{2}]^{-1/2}\exp(-\frac{[(E_{A_{3}}^{0}+\mathcal{E}_{b}^{A_{3}})-(E_{A_{1}%
}^{0}+\mathcal{E}_{b}^{A_{1}})]^{2}}{\sum_{\alpha}(\theta_{\alpha}^{A_{3}%
}-\theta_{\alpha}^{A_{1}})^{2}\hbar^{2}\omega_{\alpha}^{2}\text{csch}%
\frac{\beta\hbar\omega_{\alpha}}{2}})
\]
It can be viewed as a generalization of Holstein's result for small polaron
hopping (cf. Eq.(77) of ref.\cite{Hol2}) to the transition between two
localized states. The transfer integral $J_{A_{3}A_{1}}\varpropto e^{-qR_{31}%
}$, $q$ is some constant. In the electric conduction, only the jumps between
the nearest localized states are important. The dc conductivity from the
localized electrons is\cite{rei}
\begin{equation}
\sigma_{\text{dc}}^{LL}=\mu_{\text{hop}}^{LL}N_{L}c(1-c)e,\text{ \ \ }%
\mu_{\text{hop}}^{LL}=\frac{eR_{31}^{2}}{k_{B}T}W_{T} \label{dc}%
\end{equation}
where $n_{L}$ is the number of electrons in the localized states,
$c=n_{L}/N_{L}$. Eq.(\ref{dc}) was derived from the Kubo formula for
optical-phonon-assisted hopping\cite{Emin75}.

At `very' high temperature $k_{B}T\gtrsim2.5\hbar\overline{\omega}$, one can
use $\tanh x\thickapprox x$ and csch$x\thickapprox\frac{1}{x}$ (the error is
less than $0.003$ when $x<0.2$). Eq.(\ref{go}) is changed into%
\begin{equation}
W_{T}(A_{1}\rightarrow A_{3})=\nu_{LL}e^{-E_{a}^{LL}/k_{B}T},\text{ \ }%
\nu_{LL}=\frac{J_{A_{3}A_{1}}^{2}}{\hbar}[\frac{\pi}{\lambda_{LL}k_{B}%
T}]^{1/2}\text{,\ \ }E_{a}^{LL}=\frac{\lambda_{LL}}{4}(1+\frac{\Delta
G_{LL}^{0}}{\lambda_{LL}})^{2} \label{Mar}%
\end{equation}
where%
\begin{equation}
\Delta G_{LL}^{0}=(E_{A_{3}}^{0}+\mathcal{E}_{b}^{A_{3}})-(E_{A_{1}}%
^{0}+\mathcal{E}_{b}^{A_{1}}),\text{ \ }\lambda_{LL}=\frac{1}{2}\sum_{\alpha
}M_{\alpha}\omega_{\alpha}^{2}(\Theta_{\alpha}^{A_{3}}-\Theta_{\alpha}^{A_{1}%
})^{2} \label{de}%
\end{equation}
\ \ $\Delta G_{LL}^{0}$ is the energy difference between two localized states.
$\lambda_{LL}$ is the reorganization energy which depends on the vibrational
configurations $\{\Theta_{\alpha}^{A_{3}}\}$ and $\{\Theta_{\alpha}^{A_{1}}\}$
of the two localized states. Because $\Theta_{\alpha}^{A}$ does not have
determined sign for different state and mode, one can only roughly estimate
$\lambda_{LL}\thicksim k^{-1}(\frac{Z^{\ast}e^{2}}{4\pi\epsilon_{0}\xi^{2}%
})^{2}$. From Eq.(\ref{Mar}), we know that
$E_{a}^{LL}$ is about several tens meV. It is in agreement with observed value
of a-Si. According to Eqs.(\ref{sub}), (\ref{stad}) and (\ref{shif}), if two
localized states are in two different spatial regions, then the atoms
displaced by e-ph interaction are different for two states. $\lambda_{LL}$ can
not be zero. If the spatial regions for two localized states have overlap,
there is possibility that some atoms are moved in similar way in two states by
e-ph interaction, $\lambda_{LL}$ could be small. $\lambda_{LL}$ is zero for
two extended states, because for extended states the static displacements of
the atoms are zero. Molecular dynamics gives phonon spectrum and the
eigenvector $\Delta_{p_{A_{1}}\alpha}$ of each mode, and therefore the
reorganization energy $\lambda_{LL}$. Eq.(\ref{Mar}) is the transition
probability between two localized states in an amorphous solid, it has the
same form as Eq.(\ref{ea}).

Fig.\ref{fig1} is the temperature dependence of mobility of a-Si for
T$>250$K($\thickapprox$0.02eV), plotted from Eqs.(\ref{dc}) and (\ref{Mar})
according to a group of typical parameters: $R_{A_{3}A_{1}}=5$\AA $,$
$\lambda_{LL}$\bigskip$=0.2$eV, $\Delta G_{LL}=0.05$eV (about one half
mobility edge) and $J_{A_{3}A_{1}}=0.02$eV. Using above parameters,
$W_{T}\thicksim10^{12}$sec$^{-1}$. The computed mobility agrees with the
measurements\cite{mo,tie,cer,koc}.

\subsection{Low temperatures}

At low temperature, the probability of diagonal transition and the probability
of non-diagonal transition is comparable. The probability of diagonal
transition is given by%
\begin{equation}
W_{T}^{(d)}(A_{1}\cdots N_{\alpha}\cdots\rightarrow A_{3}\cdots N_{\alpha
}\cdots)=\frac{2\pi}{\hbar}J_{A_{3}A_{1}}^{2}\delta\lbrack(E_{A_{3}}%
^{0}+\mathcal{E}_{b}^{A_{3}})-(E_{A_{1}}^{0}+\mathcal{E}_{b}^{A_{1}})]
\label{low}%
\end{equation}%
\[
\exp\{-\sum_{\alpha}(N_{\alpha}+\frac{1}{2})(\theta_{\alpha}^{A_{3}}%
-\theta_{\alpha}^{A_{1}})^{2}\}
\]
in the same approximation as Eq.(\ref{sep}). The first line is the expression
for the transition between two \textit{electronic} states without the phonon
environment. The last exponential factor of Eq.(\ref{low}) also appears in the
corresponding expression for non-diagonal transition. It can be explained as
the decrease of transfer integral caused by electron-vibration coupling.

At zero temperature $N_{\alpha}=0$, and only zero point vibration remains. The
transition between two localized states is then pure quantum tunneling. At
finite temperature, one needs to average Eq.(\ref{low}) over the equilibrium
phonon distribution:%
\begin{equation}
W_{T}^{(d)}(A_{1}\rightarrow A_{3})=\frac{2\pi}{\hbar}J_{A_{3}A_{1}}^{2}%
\delta\lbrack(E^{A_{1}}+\mathcal{E}_{b}^{A_{1}})-(E^{A_{3}}+\mathcal{E}%
_{b}^{A_{3}})] \label{dL}%
\end{equation}%
\[
\exp\{-\frac{1}{2}\sum_{\alpha}(\theta_{\alpha}^{A_{3}}-\theta_{\alpha}%
^{A_{1}})^{2}\coth\frac{\beta\hbar\omega_{\alpha}}{2}\}
\]
The lower value of diagonal transition probability in non-zero temperature
comparing with that of zero temperature may be understood as that thermal
vibrations disturb quantum tunneling. Group the exponential function and
$J_{A_{3}A_{1}}^{2}$ in Eq.(\ref{dL}) together, one may say that transfer
integral is reduced in a multi-phonon process. This is similar to that the
effective band width decreases with e-ph coupling which was derived from a
different model\cite{cap1,fei}.

\begin{figure}[h]
\begin{center}
\resizebox{90mm}{!}{\includegraphics{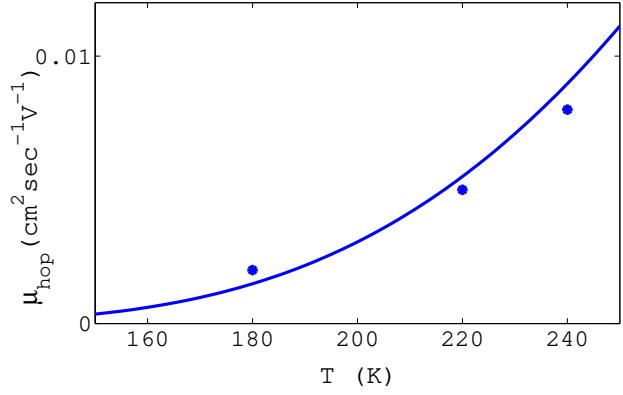}}
\end{center}
\par
.\caption{(Color online) Mobility from LL transition in a-Si (T$<$ 250K):
solid line is computed from Eq.(\ref{di}) with parameters: $\Delta
G_{LL}=0.05eV,~J_{LL}=0.02eV,~\bar{\omega}=2\pi\times850$cm$^{-1}$. The star
symbols are experimental data\cite{mo}.}%
\label{fig2}%
\end{figure}

In low temperature $k_{B}T\lesssim\hbar\overline{\omega}/10$ (csch$x<0.0135$
when $x>5$), the exponent in the last exponential of Eq.(\ref{TT}) is small.
The exponential\ can be expanded in Taylor series of its exponent, then the
`time' integral can be finished. Denote%
\begin{equation}
f(\omega_{\alpha})=\frac{1}{2}(\theta_{\alpha}^{A_{3}}-\theta_{\alpha}^{A_{1}%
})^{2}\text{csch}\frac{\beta\hbar\omega_{\alpha}}{2} \label{f}%
\end{equation}
the result is%
\begin{equation}
W_{T}(A_{1}\rightarrow A_{3})=\frac{2\pi J_{A_{3}A_{1}}^{2}}{\hbar}\exp
\{\frac{-\beta\Delta G_{LL}^{0}}{2}\}\exp\{-\frac{1}{2}\sum_{\alpha}%
(\theta_{\alpha}^{A_{3}}-\theta_{\alpha}^{A_{1}})^{2}\coth\frac{^{\beta
\hbar\omega_{\alpha}}}{2}\} \label{di}%
\end{equation}%
\[
\{\sum_{\alpha}f(\omega_{\alpha})\frac{1}{2}[\delta(\Delta G_{LL}^{0}%
+\hbar\omega_{\alpha})+\delta(\Delta G_{LL}^{0}-\hbar\omega_{\alpha})]
\]%
\begin{align*}
&  +\sum_{\alpha\alpha^{\prime}}f(\omega_{\alpha})f(\omega_{\alpha^{\prime}%
})\frac{1}{8}[\delta(\Delta G_{LL}^{0}+\hbar\omega_{\alpha}+\hbar
\omega_{\alpha^{\prime}})+\delta(\Delta G_{LL}^{0}-\hbar\omega_{\alpha}%
-\hbar\omega_{\alpha^{\prime}})\\
&  +\delta(\Delta G_{LL}^{0}+\hbar\omega_{\alpha}-\hbar\omega_{\alpha^{\prime
}})+\delta(\Delta G_{LL}^{0}-\hbar\omega_{\alpha}+\hbar\omega_{\alpha^{\prime
}})]\\
&  +\cdots\}
\end{align*}
In fact $\coth\frac{^{\beta\hbar\omega_{\alpha}}}{2}$ in the second
exponential factor can be replaced by 1 when $k_{B}T\lesssim\hbar
\overline{\omega}/8$ (coth$x-1<0.0007$ when $x>4$). The terms in the curly
bracket are single-phonon transition, 2-phonon transition, etc. $f$ quickly
decreases with decreasing temperature, only single-phonon processes are
important at low temperature. Because low frequency modes have larger density
of states in amorphous solids, multi-phonon processes are also visible at low
temperature. Fig.\ref{fig2} is the mobility from LL transition at low
temperature regime calculated from Eq.(\ref{di}) with parameters $\Delta
G_{LL}=0.05eV,~J_{LL}=0.02eV,~\bar{\omega}=2\pi\times850$cm$^{-1}$ (taken from
ref.\onlinecite{col}). No reorganization is needed at low temperature regime,
but function f decreases with temperature rapidly. The separation temperature
T=250K in data fitting\cite{mo,cer} is higher than the estimated upper limit
of low temperature regime $\bar{\omega}/10\sim122$K.

The activation energy for the transition between two mid-gap states is just
the energy difference between the two levels. For mid-gap states, variable
range hopping is an effective conduction mechanism at low
temerature\cite{motv,aps}. Eq.(\ref{di}) is the conduction of tail states at
low temperature due to non-diagonal LL transition. The activation energy
$\frac{\Delta G_{LL}^{0}}{2}$ is one half the energy difference between two
localized states. This is contrast with the high temperature partner
Eq.(\ref{Mar}) where activation energy involves reorganization energy.
Therefore variable range transition is also an effective conduction mechanism
for tail states as non-diagonal transition in the low temperature regime
($k_{B}T\lesssim\hbar\overline{\omega}/10$). Thus $\frac{\hbar\overline
{\omega}}{10k_{B}}$ is a rough upper limit temperature of variable range
\ hopping. In a-Si, $\overline{\omega}/10\thicksim122$K roughly agrees with
the upper limit temperature\cite{md} of variable range hopping (100K).

\section{transition from a localized state to an extended state}

If at the initial moment ($t=0$) the \textquotedblleft one electron + many
nuclei\textquotedblright\ system is at state $|A_{1}\cdots N_{\alpha}^{\prime
}\cdots\rangle$, then $C^{A_{1}}(\cdots N_{\alpha}^{\prime}\cdots;t=0)=1$ and
other coefficients vanish. At later moment $t$, the probability amplitude
$F^{B_{2}}(\cdots N_{\alpha}\cdots;t)$ is simply determined by the first order
perturbation approximation of Eq.(\ref{fq}):%

\begin{equation}
i\hbar\frac{\partial F_{\{N_{\alpha}\}}^{B_{2}}}{\partial t}=\langle
B_{2}\cdots N_{\alpha}\cdots|V_{LE}^{tr}|A_{1}\cdots N_{\alpha}^{\prime}%
\cdots\rangle C_{\{N_{\alpha}^{\prime}\}}^{A_{1}}e^{it(\mathcal{E}_{B_{2}%
}^{\{N_{\alpha}\}}-\mathcal{E}_{A_{1}}^{\{N_{\alpha}^{\prime}\}})/\hbar}
\label{PLET}%
\end{equation}
It describes the transition from localized state $|A_{1}\cdots N_{\alpha
}^{\prime}\cdots\rangle$ to extended state $|B_{2}\cdots N_{\alpha}%
\cdots\rangle$ caused by the transfer integral $J^{\prime}{}_{B_{2}A_{1}}$.
Expand $\Phi_{N_{\alpha}^{\prime}}(\theta_{\alpha}-\theta_{\alpha}^{A_{1}})$
around $\theta_{\alpha}$ to the second order of $\theta_{\alpha}^{A_{1}}$, the
integral in matrix element (\ref{LET}) can be effected. Then the transition
probability from localized state $A_{1}$ to extended state $B_{2}$ is%
\[
W_{T}(A_{1}\rightarrow B_{2})=\frac{J^{\prime2}{}_{B_{2}A_{1}}}{\hbar^{2}%
}e^{-\beta(E_{B_{2}}-E_{A_{1}}^{0}-\mathcal{E}_{A_{1}}^{b})/2}\exp\{-\frac
{1}{2}\sum_{\alpha}(\theta_{\alpha}^{A_{1}})^{2}\coth\frac{\beta\hbar
\omega_{\alpha}}{2}\}
\]%
\begin{equation}
\int_{-t}^{t}d\tau\exp\{\frac{i\tau}{\hbar}(E_{B_{2}}-E_{A_{1}}^{0}%
-\mathcal{E}_{A_{1}}^{b})\}[\exp\{\frac{1}{2}\sum_{\alpha}(\theta_{\alpha
}^{A_{1}})^{2}\text{csch}\frac{\beta\hbar\omega_{\alpha}}{2}\cos\omega
_{\alpha}\tau\}-1] \label{ple4}%
\end{equation}

For moderately high temperature $k_{B}T\gtrsim\hbar\overline{\omega}$, the 1
in the last square bracket can be neglected. Expand $\cos\omega_{\alpha}%
\tau\thickapprox1-\frac{\tau^{2}\omega_{\alpha}^{2}}{2}$, the integral is a
Gaussian integral,%
\[
W_{T}^{LE}=\frac{J^{\prime2}{}_{B_{2}A_{1}}}{\hbar^{2}}e^{-\beta(E_{B_{2}%
}-E_{A_{1}}^{0}-\mathcal{E}_{A_{1}}^{b})/2}\exp\{-\frac{1}{2}\sum_{\alpha
}(\theta_{\alpha}^{A_{1}})^{2}\tanh\frac{\beta\hbar\omega_{\alpha}}{4}\}
\]%
\begin{equation}
(2\pi)^{1/2}[\frac{1}{2}\sum_{\alpha}(\theta_{\alpha}^{A_{1}})^{2}%
\omega_{\alpha}^{2}\text{csch}\frac{\beta\hbar\omega_{\alpha}}{2}]^{-1/2}%
\exp(-\frac{[(E_{A_{3}}^{0}+\mathcal{E}_{b}^{A_{3}})-(E_{A_{1}}^{0}%
+\mathcal{E}_{b}^{A_{1}})]^{2}}{\sum_{\alpha}(\theta_{\alpha}^{A_{1}}%
)^{2}\hbar^{2}\omega_{\alpha}^{2}\text{csch}\frac{\beta\hbar\omega_{\alpha}%
}{2}}) \label{ple5}%
\end{equation}
For `very' high temperature $k_{B}T\gtrsim2.5\hbar\overline{\omega},$
Eq.(\ref{ple5}) is further simplified as:%
\begin{equation}
W_{T}^{LE}=\nu_{LE}e^{-E_{a}^{LE}/k_{B}T},\text{ \ \ }E_{a}^{LE}=\frac
{\lambda_{LE}}{4}(1+\frac{\Delta G_{LE}^{0}}{\lambda_{LE}})^{2} \label{LEmar}%
\end{equation}
where%
\begin{equation}
\Delta G_{LE}^{0}=E_{B_{2}}-(E_{A_{1}}^{0}+\mathcal{E}_{A_{1}}^{b})
\label{ELd}%
\end{equation}
is the energy difference between extended state $B_{2}$ and localized state
$A_{1}$.
\begin{equation}
\nu_{LE}=\frac{J^{\prime2}{}_{B_{2}A_{1}}}{\hbar}(\frac{\pi}{\lambda_{LE}%
k_{B}T})^{1/2},\text{ }\lambda_{LE}=\frac{1}{2}\sum_{\alpha}M_{\alpha}%
\omega_{\alpha}^{2}(\Theta_{\alpha}^{A_{1}})^{2} \label{rLE}%
\end{equation}
$\lambda_{LE}$ is the reorganization energy for transition from localized
state $A_{1}$ to extended state $B_{2}$. It is interesting to notice that
activation energy $E_{a}^{LE}$ for LE transition can be obtained by assume
$\Theta_{\alpha}^{A_{3}}=0$ in $\lambda_{LL}$. Transition from a localized
state to an extended state corresponds to that a particle escapes a barrier
along reaction path\cite{kram}.

From their expressions (\ref{de}) and (\ref{rLE}), we know $\lambda_{LE}$ is
same order of magnitude as $\lambda_{LL}$. $\Delta G_{LE}^{0}$ is order of
mobility edge, several times larger than $\Delta G_{LL}^{0}$. $J^{\prime}%
{}_{B_{2}A_{1}}$ is several times larger than $J^{\prime}{}_{A_{3}A_{1}}$. The
spatial displacement of the electron in a LE transition is about the linear
size of the localized state. In general the LE transition probability is
smaller than that of the LL transition. From Eqs.(\ref{LEmar}) and
(\ref{Mar}), $W_{T}^{LE}$ becomes comparable with $W_{T}^{LL}$ only when
temperature is higher than $k_{B}^{-1}(E_{a}^{LE}-E_{a}^{LL})[2\ln
(J_{LE}/J_{LL})]^{-1}$. The mobility edge of a-Si is about 0.1eV, therefore LL
transition is dominant in intrinsic a-Si below 580K. However if
\textit{higher} localized states close to the mobility edge are occupied due
to doping, there exist some extended states which satisfy $\Delta G_{LE}%
^{0}\thicksim\Delta G_{LL}^{0}$. For these LE transitions, $E_{a}^{LE}$ is
comparable to $E_{a}^{LL}$. The LE transition probability is about 10 times
larger than that of LL transition. For these higher localized states, using
parameters given for LL$\ $transition in a-Si, $W_{T}^{LE}\thicksim10^{13}%
$Sec$^{-1}$. Comparing Eqs.(\ref{dc}) and (\ref{rLE}), from the value of
hopping mobility 0.05 cm$^{2}$Sec$^{-1}$V$^{-1}$, we expect the mobility from
LE transition is several tenths cm$^{2}$Sec$^{-1}$V$^{-1}$ for a-Si, the same
order of magnitude as observed `drift mobility'\cite{tie,cer}.

The probability of diagonal transition is given by%
\begin{equation}
W_{T}^{(d)}(B_{2}\cdots N_{\alpha}\cdots\rightarrow A_{1}\cdots N_{\alpha
}\cdots)=\frac{2\pi}{\hbar}J^{\prime2}{}_{B_{2}A_{1}}\delta\lbrack E_{B_{2}%
}-(E_{A_{1}}^{0}+\mathcal{E}_{A_{1}}^{b})] \label{dui}%
\end{equation}%
\[
\exp\{-\sum_{\alpha}(N_{\alpha}+\frac{1}{2})(\theta_{\alpha}^{A_{1}})^{2}\}
\]
The last exponential factor can be explained as the decrease of transition
integral caused by electron-vibration coupling. At zero temperature
$N_{\alpha}=0$, the transition from a localized state to an extended state is
then pure quantum tunneling. At finite temperature, one needs to average
Eq.(\ref{dui}) over the equilibrium phonon distribution:%
\begin{equation}
W_{T}^{(d)}(A_{1}\rightarrow B_{2})=\frac{2\pi}{\hbar}J^{\prime2}{}%
_{B_{2}A_{1}}\exp\{-\frac{1}{2}\sum_{\alpha}(\theta_{\alpha}^{A_{1}})^{2}%
\coth\frac{\beta\hbar\omega_{\alpha}}{2}\}\delta\lbrack E_{B_{2}}-(E_{A_{1}%
}^{0}+\mathcal{E}_{b}^{A_{1}})] \label{findi}%
\end{equation}

In low temperature $k_{B}T\lesssim\hbar\overline{\omega}/10$, the exponent in
the last exponential of Eq.(\ref{ple4}) is small. The exponential\ can be
expanded in Taylor series of its exponent, then the `time' integral can be
finished. Denote%
\begin{equation}
f_{LE}(\omega_{\alpha})=\frac{1}{2}(\theta_{\alpha}^{A_{1}})^{2}%
\text{csch}\frac{\beta\hbar\omega_{\alpha}}{2} \label{fle}%
\end{equation}
the result is%
\begin{equation}
W_{T}(A_{1}\rightarrow B_{2})=\frac{2\pi J^{\prime2}{}_{B_{2}A_{1}}}{\hbar
}\exp\{\frac{-\beta\Delta G_{LE}^{0}}{2}\}\exp\{-\frac{1}{2}\sum_{\alpha
}(\theta_{\alpha}^{A_{1}})^{2}\coth\frac{^{\beta\hbar\omega_{\alpha}}}{2}\}
\label{nond}%
\end{equation}%
\[
\{\sum_{\alpha}f_{LE}(\omega_{\alpha})\frac{1}{2}[\delta(\Delta G_{LE}%
^{0}+\hbar\omega_{\alpha})+\delta(\Delta G_{LE}^{0}-\hbar\omega_{\alpha})]
\]%
\begin{align*}
&  +\sum_{\alpha\alpha^{\prime}}f_{LE}(\omega_{\alpha})f_{LE}(\omega
_{\alpha^{\prime}})\frac{1}{8}[\delta(\Delta G_{LE}^{0}+\hbar\omega_{\alpha
}+\hbar\omega_{\alpha^{\prime}})+\delta(\Delta G_{LE}^{0}-\hbar\omega_{\alpha
}-\hbar\omega_{\alpha^{\prime}})\\
&  +\delta(\Delta G_{LE}^{0}+\hbar\omega_{\alpha}-\hbar\omega_{\alpha^{\prime
}})+\delta(\Delta G_{LE}^{0}-\hbar\omega_{\alpha}+\hbar\omega_{\alpha^{\prime
}})]\\
&  +\cdots\}
\end{align*}

Formally type (2) transition is quite similar to type (1) transition. To
obtain the former, one needs to make substitutions: $J{}_{A_{3}A_{1}%
}\rightarrow J^{\prime}{}_{B_{2}A_{1}}$ $\Theta_{\alpha}^{A_{3}}%
-\Theta_{\alpha}^{A_{1}}\rightarrow\Theta_{\alpha}^{A_{1}}$ and $[(E^{A_{1}%
}+\mathcal{E}_{b}^{A_{1}})-(E^{A_{3}}+\mathcal{E}_{b}^{A_{3}})]\rightarrow
\lbrack E_{B_{2}}-(E_{A_{1}}^{0}+\mathcal{E}_{b}^{A_{1}})]$ in corresponding
expressions of the former. However the physical meaning of the two are
completely different: they are reflected in the transfer integrals
Eqs.(\ref{tra1}) and (\ref{tra2}).

Consider a localized state and an extended state, both of them are close to
the mobility edge. Then\ $\Delta G_{LE}^{0}$ is small. For such a localized
state, the displacements of the atoms induced by the polarization of the
localized state (cf. $\Theta_{\alpha}^{A_{1}}$ in Eq.(\ref{or})) are small.
Its atomic configuration is similar to that of an extended state. Therefore
the reorganization energy $\lambda_{LE}$ is small. The transition probability
(\ref{LEmar}) can be large. Similar picture has been suggested long time ago
in name of phonon-induced delocalization\cite{kik}. The inelastic process
makes the concept of localization meaningless\cite{thou,im} for the states
close to the mobility edge.

\section{transition from an extended state to a localized state}

If at the initial moment the electron+nuclei system is at state $|B_{1}\cdots
N_{\alpha}^{\prime}\cdots\rangle$, then $F^{B_{1}}(\cdots N_{\alpha}^{\prime
}\cdots;t=0)=1$ and other coefficients are zero. At later moment $t$,
probability amplitude $C^{A_{2}}(\cdots N_{\alpha}\cdots;t)$ is determined by
first order perturbation theory from Eq.(\ref{cq}):%
\begin{equation}
i\hbar\frac{\partial C^{A_{2}}}{\partial t}=\langle A_{2}\cdots N_{\alpha
}\cdots|V_{EL}^{e-ph}|B_{1}\cdots N_{\alpha}^{\prime}\cdots\rangle F^{B_{1}%
}(\cdots N_{\alpha}^{\prime}\cdots;t)e^{it(\mathcal{E}_{A_{2}}^{\{N_{\alpha
}\}}-\mathcal{E}_{B_{1}}^{\{N_{\alpha}^{\prime}\}})/\hbar} \label{ELP}%
\end{equation}

Expand $\Phi_{N_{\alpha}}(\theta_{\alpha}-\theta_{\alpha}^{A_{2}})$ around
$\theta_{\alpha}$ to second order of $\theta_{\alpha}^{A_{2}}$, effect the
multiple integral over normal modes, the matrix element (\ref{ELT}) can be
calculated. The diagonal transition probability (in which the phonon state
invariant before and after the change in electronic state) induced by
electron-phonon interaction is
\begin{equation}
W_{T}^{(d)}(B_{1}\rightarrow A_{2})=\delta(E_{A_{2}}^{0}+\mathcal{E}_{A_{2}%
}^{b}-E_{B_{1}})\frac{2\pi}{\hbar}\exp\{-\frac{1}{2}\sum_{\alpha}%
(\theta_{\alpha}^{A_{2}})^{2}\coth\frac{\beta\hbar\omega_{\alpha}}{2}%
\}\frac{1}{4}\sum_{\alpha^{\prime}}(K_{A_{2}B_{1}}^{\prime\alpha}%
\Theta_{\alpha^{\prime}}^{A_{2}})^{2} \label{eld2}%
\end{equation}
The probability of the transition from extended state $|B_{1}\rangle$ to
localized state $|A_{2}\rangle$ is%
\begin{equation}
W_{T}(B_{1}\rightarrow A_{2})=\frac{1}{\hbar^{2}}\exp\{-\frac{\beta}%
{2}(E_{A_{2}}^{0}+\mathcal{E}_{A_{2}}^{b}-E_{B_{1}})\}\exp\{-\frac{1}{2}%
\sum_{\alpha}(\theta_{\alpha}^{A_{2}})^{2}\coth\frac{\beta\hbar\omega_{\alpha
}}{2}\}(I_{1}+I_{2}) \label{el6}%
\end{equation}
where%
\begin{equation}
I_{1}=\int_{-t}^{t}d\tau\exp\{\frac{i\tau}{\hbar}(E_{A_{2}}^{0}+\mathcal{E}%
_{A_{2}}^{b}-E_{B_{1}})\}\exp\{\frac{1}{2}\sum_{\alpha}(\theta_{\alpha}%
^{A_{2}})^{2}\text{csch}\frac{\beta\hbar\omega_{\alpha}}{2}\cos\omega_{\alpha
}\tau\} \label{i1}%
\end{equation}%
\[
\lbrack\sum_{\alpha^{\prime}}(\frac{1}{2}(K_{A_{2}B_{1}}^{\prime\alpha
^{\prime}})^{2}+\frac{1}{4}(K_{A_{2}B_{1}}^{\prime\alpha^{\prime}}%
\theta_{\alpha^{\prime}}^{A_{2}})^{2}\coth\frac{\beta\hbar\omega
_{\alpha^{\prime}}}{2})\frac{\hbar}{M_{\alpha^{\prime}}\omega_{\alpha^{\prime
}}}\text{csch}\frac{\beta\hbar\omega_{\alpha^{\prime}}}{2}\cos\omega
_{\alpha^{\prime}}\tau
\]%
\[
-\frac{1}{4}\sum_{\alpha^{\prime}}\frac{\hbar}{M_{\alpha^{\prime}}%
\omega_{\alpha^{\prime}}}(K_{A_{2}B_{1}}^{\prime\alpha^{\prime}}\theta
_{\alpha^{\prime}}^{A_{2}})^{2}\text{csch}^{2}\frac{\beta\hbar\omega
_{\alpha^{\prime}}}{2}\cos^{2}\omega_{\alpha^{\prime}}\tau]
\]
and%
\begin{equation}
I_{2}=[\frac{1}{4}\sum_{\alpha^{\prime}}(K_{A_{2}B_{1}}^{\prime\alpha}%
\Theta_{\alpha^{\prime}}^{A_{2}})^{2}]\int_{-t}^{t}d\tau\exp\{\frac{i\tau
}{\hbar}(E_{A_{2}}^{0}+\mathcal{E}_{A_{2}}^{b}-E_{B_{1}})\}[\exp\{\frac{1}%
{2}\sum_{\alpha}(\theta_{\alpha}^{A_{2}})^{2}\text{csch}\frac{\beta\hbar
\omega_{\alpha}}{2}\cos\omega_{\alpha}\tau\}-1] \label{i2}%
\end{equation}

At moderately high temperature $k_{B}T\gtrsim\hbar\overline{\omega}$, by
expanding $\cos x\thickapprox1-\frac{x^{2}}{2}$, the integrals $I_{1}$ and
$I_{2}$ can be carried out. One has
\begin{equation}
W_{T}^{EL}(B_{1}\rightarrow A_{2})=\frac{1}{\hbar^{2}}\exp\{-\frac{\beta}%
{2}(E_{A_{2}}^{0}+\mathcal{E}_{A_{2}}^{b}-E_{B_{1}})\}\exp\{-\frac{1}{2}%
\sum_{\alpha}(\theta_{\alpha}^{A_{2}})^{2}\tanh\frac{\beta\hbar\omega_{\alpha
}}{4}\} \label{el7}%
\end{equation}%
\[
(2\pi)^{1/2}[\frac{1}{2}\sum_{\alpha}(\theta_{\alpha}^{A_{2}})^{2}%
\omega_{\alpha}^{2}\text{csch}\frac{\beta\hbar\omega_{\alpha}}{2}]^{-1/2}%
\exp(-\frac{(E_{A_{2}}^{0}+\mathcal{E}_{A_{2}}^{b}-E_{B_{1}})^{2}}%
{\sum_{\alpha}(\theta_{\alpha}^{A_{2}})^{2}\hbar^{2}\omega_{\alpha}%
^{2}\text{csch}\frac{\beta\hbar\omega_{\alpha}}{2}})
\]%
\[
\{\frac{1}{4}\sum_{\alpha^{\prime}}(K_{A_{2}B_{1}}^{\prime\alpha}%
\Theta_{\alpha^{\prime}}^{A_{2}})^{2}+\sum_{\alpha^{\prime}}[\frac
{(K_{A_{2}B_{1}}^{\prime\alpha})^{2}\hbar}{2M_{\alpha^{\prime}}\omega
_{\alpha^{\prime}}}\text{csch}\frac{\beta\hbar\omega_{\alpha^{\prime}}}%
{2}+\frac{(K_{A_{2}B_{1}}^{\prime\alpha}\Theta_{\alpha^{\prime}}^{A_{2}})^{2}%
}{8}\text{sech}^{2}\frac{\beta\hbar\omega_{\alpha^{\prime}}}{4}]
\]%
\[
-[2(\sum_{\alpha}\omega_{\alpha}^{2}(\theta_{\alpha}^{A_{2}})^{2}%
\text{csch}\frac{\beta\hbar\omega_{\alpha}}{2})^{-1}-4\frac{(E_{A_{2}}%
^{0}+\mathcal{E}_{A_{2}}^{b}-E_{B_{1}})^{2}}{(\sum_{\alpha}\hbar\omega
_{\alpha}^{2}(\theta_{\alpha}^{A_{2}})^{2}\text{csch}\frac{\beta\hbar
\omega_{\alpha}}{2})^{2}}]
\]%
\[
\times\sum_{\alpha^{\prime}}[\frac{(K_{A_{2}B_{1}}^{\prime\alpha})^{2}%
\hbar\omega_{\alpha^{\prime}}}{4M_{\alpha^{\prime}}}\text{csch}\frac
{\beta\hbar\omega_{\alpha^{\prime}}}{2}+\frac{\omega_{\alpha^{\prime}}^{2}}%
{8}(K_{A_{2}B_{1}}^{\prime\alpha}\Theta_{\alpha^{\prime}}^{A_{2}})^{2}%
\coth\frac{\beta\hbar\omega_{\alpha^{\prime}}}{2}\text{csch}\frac{\beta
\hbar\omega_{\alpha^{\prime}}}{2}-\frac{\omega_{\alpha^{\prime}}^{2}%
(K_{A_{2}B_{1}}^{\prime\alpha}\Theta_{\alpha^{\prime}}^{A_{2}})^{2}}%
{4}\text{csch}^{2}\frac{\beta\hbar\omega_{\alpha}}{2}]\}
\]
It is interesting to notice that the first two lines of Eq.(\ref{el7}) have
the\ same structure as Eq.(\ref{go})\ and Eq.(\ref{ple5}). The non-radiative
lifetime of an electron in an extended state is defined as $t_{\text{life}%
}=1/\sum_{A_{2}}W_{T}^{EL}$.

For very high temperature $k_{B}T\gtrsim2.5\hbar\overline{\omega}$,
Eq.(\ref{el7}) becomes%
\begin{equation}
W_{EL}=\nu_{EL}e^{-E_{a}^{EL}/k_{B}T} \label{elp}%
\end{equation}
with%
\begin{equation}
\nu_{EL}=\frac{1}{\hbar}(\frac{\pi}{k_{B}T\lambda_{EL}})^{1/2}\{\frac{1}%
{4}\sum_{\alpha^{\prime}}(K_{A_{2}B_{1}}^{\prime\alpha^{\prime}}\Theta
_{\alpha^{\prime}}^{A_{2}})^{2}+\sum_{\alpha^{\prime}}[\frac{(K_{A_{2}B_{1}%
}^{\prime\alpha^{\prime}})^{2}\hbar}{2M_{\alpha^{\prime}}\omega_{\alpha
^{\prime}}}\text{csch}\frac{\beta\hbar\omega_{\alpha^{\prime}}}{2}%
+\frac{(K_{A_{2}B_{1}}^{\prime\alpha}\Theta_{\alpha^{\prime}}^{A_{2}})^{2}}%
{8}\text{sech}^{2}\frac{\beta\hbar\omega_{\alpha^{\prime}}}{4}] \label{nuel}%
\end{equation}%
\[
-[2(\sum_{\alpha}\omega_{\alpha}^{2}(\theta_{\alpha}^{A_{2}})^{2}%
\text{csch}\frac{\beta\hbar\omega_{\alpha}}{2})^{-1}-4\frac{(E_{A_{2}}%
^{0}+\mathcal{E}_{A_{2}}^{b}-E_{B_{1}})^{2}}{(\sum_{\alpha}\hbar\omega
_{\alpha}^{2}(\theta_{\alpha}^{A_{2}})^{2}\text{csch}\frac{\beta\hbar
\omega_{\alpha}}{2})^{2}}]\times
\]%
\[
\times\sum_{\alpha^{\prime}}[\frac{(K_{A_{2}B_{1}}^{\prime\alpha^{\prime}%
})^{2}\hbar\omega_{\alpha^{\prime}}}{4M_{\alpha^{\prime}}}\text{csch}%
\frac{\beta\hbar\omega_{\alpha^{\prime}}}{2}+\frac{\omega_{\alpha^{\prime}%
}^{2}(K_{A_{2}B_{1}}^{\prime\alpha^{\prime}}\Theta_{\alpha^{\prime}}^{A_{2}%
})^{2}\coth\frac{\beta\hbar\omega_{\alpha^{\prime}}}{2}}{8}\text{csch}%
\frac{\beta\hbar\omega_{\alpha^{\prime}}}{2}-\frac{\omega_{\alpha^{\prime}%
}^{2}(K_{A_{2}B_{1}}^{\prime\alpha^{\prime}}\Theta_{\alpha^{\prime}}^{A_{2}%
})^{2}}{4}\text{csch}^{2}\frac{\beta\hbar\omega_{\alpha^{\prime}}}{2}]\}
\]
and%
\begin{equation}
E_{a}^{EL}=\frac{\lambda_{EL}}{4}(1+\frac{\Delta G_{EL}^{0}}{\lambda_{EL}%
})^{2},\text{ \ \ }\lambda_{EL}=\frac{1}{2}\sum_{\alpha}M_{\alpha}%
\omega_{\alpha}^{2}(\Theta_{\alpha}^{A_{2}})^{2},\text{\ \ \ }\Delta
G_{EL}^{0}=E_{A_{2}}^{0}+\mathcal{E}_{A_{2}}^{b}-E_{B_{1}}<0 \label{acel}%
\end{equation}

$\lambda_{EL}$ is the same order of magnitude as $\lambda_{LL}$. $\Delta
G_{EL}^{0}<0$, $E_{a}^{EL}$ is smaller than $E_{a}^{LL}$. $K_{A_{2}B_{1}%
}^{\prime\alpha^{\prime}}\Theta_{\alpha^{\prime}}^{A_{2}}$ is the same order
magnitude as $J_{A_{3}A_{1}}$, EL transition probability is larger than that
of LL transition. With the parameters for LL$\ $transition in a-Si,
$W_{T}^{EL}\thicksim10^{13}-10^{14}$sec$^{-1}$. In a-Si:H, it leads to a
mobility about several tenths of cm$^{2}$sec$^{-1}$V$^{-1}$, a value between
the observed `hopping mobility' and `drift mobility'\cite{cer,koc}. However
the population in extended states is meaningful only when temperature is high
enough or doping is heavy enough. For intrinsic semiconductor, EL transition
is not important for conduction at not too high temperature.

\begin{figure}[h]
\begin{center}
\resizebox{90mm}{!}{\includegraphics{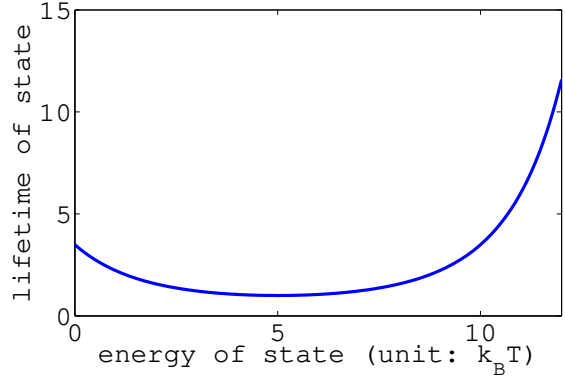}}
\end{center}
\par
.\caption{(Color online) Schematic illustration of lifetimes of states as
function of energy: energy unit is k$_{B}T$. $\lambda=5$, E$_{L}$ is taken as
energy zero, the vertical axis is the relative lifetimes of states relative to
the shortest one. Marcus inverted region: $E>8$, lifetime increases with
energy. The short lifetime belt exists in $3<E<8$. The vertical axis is scaled
by $\nu_{EL}$}%
\label{fig3}%
\end{figure}

$\Delta G_{EL}^{0}<0$ has a deep consequence. From the expression (\ref{ev})
for $\mathcal{E}_{A_{2}}^{b}$ and the order of magnitude of mobility
edge\cite{sca}, the energy difference $\Delta G_{EL}^{0}<0$ is order of
several tenths eV. For the extended states with $|\Delta G_{EL}^{0}%
|/\lambda_{EL}<1$, we are in normal regime: the higher energy of an extended
state (i.e. $\frac{\Delta G_{EL}^{0}}{\lambda_{EL}}$ more negative but still
$|\Delta G_{EL}^{0}|/\lambda_{EL}<1$), the smaller the activation energy
$E_{a}^{EL}$. The higher extended state has shorter lifetime ($0<|\Delta
G_{EL}^{0}|/k_{B}T<5$ cf. Fig.\ref{fig3}, one takes $\lambda_{EL}/k_{B}T=5$),
the electron has less time moves in extended state, it will contribute less to
electric conductivity. For extended states with energies \textit{far above}
the mobility edge (such that $|\Delta G_{EL}^{0}|/\lambda_{EL}>1$), we are in
Marcus inverted regime ($|\Delta G_{EL}^{0}|/k_{B}T>5$ cf. Fig.\ref{fig3}):
the higher the energy of an extended state, the larger the activation energy.
The higher extended states have long lifetime (as usual experience: the higher
energy of the initial state, the faster of the decay), and they will
contribute conductivity more. In the middle of the two regimes, $\frac{\Delta
G_{EL}^{0}}{\lambda_{EL}}\thickapprox-1$. For these extended states,
\textit{no activation energy is needed for the transition to localized state}.
They will quickly decay to the localized states. Reflected in
photoluminescence, the luminescence insisting time as a function of
luminescence frequency decreases with frequency. In a crystal, phonon assisted
non-radiative transition is slowed down by the energy-momentum conservation
law, the lifetime of an excited state is long. This phenomenon was found some
times ago: in c-Si/SiO2 quantum well structure, the photoluminescence lifetime
is about 1ms, the photoluminescence life time is not sensitive to the
monitored wavelength\cite{oka}. Comparing with crystalline quantum well, the
photoluminescence lifetime of a-Si/SiO$_{2}$ structure becomes shorter with a
decrease in monitored wavelength\cite{kan} 13ns at 550nm and 143ns at 750nm.
The trend is consisten with the left half part of Fig.\ref{fig3}.

The preceding discussion is also applicable to the transitions
from the extended states in conduction band to the unoccupied
\textit{localized} states in valence band.

At low temperature $k_{B}T\lesssim\hbar\overline{\omega}/10$, one can expand
the exponentials in Eqs.(\ref{i1}) and (\ref{i2}) into power series. Then the
integrals can be carried out term by term. To 2-phonon processes, the
transition probability from extended state $|B_{1}\rangle$ to localized state
$|A_{2}\rangle$ is%
\begin{equation}
W_{T}(B_{1}\rightarrow A_{2})=\frac{2\pi}{\hbar}\exp\{-\frac{\beta}%
{2}(E_{A_{2}}^{0}+\mathcal{E}_{A_{2}}^{b}-E_{B_{1}})\}\exp[-\frac{1}{2}%
\sum_{\alpha}(\theta_{\alpha}^{A_{2}})^{2}\coth\frac{\beta\hbar\omega_{\alpha
}}{2}] \label{loweL}%
\end{equation}

\[
\{\frac{1}{2}\sum_{\alpha^{\prime}}[\frac{(K_{A_{2}B_{1}}^{\prime\alpha}%
)^{2}\hbar}{2M_{\alpha^{\prime}}\omega_{\alpha^{\prime}}}+\frac{(K_{A_{2}%
B_{1}}^{\prime\alpha}\Theta_{\alpha^{\prime}}^{A_{2}})^{2}}{4}\coth\frac
{\beta\hbar\omega_{\alpha^{\prime}}}{2}]\text{csch}\frac{\beta\hbar
\omega_{\alpha^{\prime}}}{2}%
\]%
\[
\times\lbrack\delta(E_{A_{2}}^{0}+\mathcal{E}_{A_{2}}^{b}-E_{B_{1}}%
+\hbar\omega_{\alpha^{\prime}})+\delta(E_{A_{2}}^{0}+\mathcal{E}_{A_{2}}%
^{b}-E_{B_{1}}-\hbar\omega_{\alpha^{\prime}})]
\]%
\[
+\sum_{\alpha^{\prime}}\frac{(K_{A_{2}B_{1}}^{\prime\alpha}\Theta
_{\alpha^{\prime}}^{A_{2}})^{2}}{8}(1-\text{csch}^{2}\frac{\beta\hbar
\omega_{\alpha^{\prime}}}{2})\sum_{\alpha^{\prime\prime}}f_{EL}(\omega
_{\alpha^{\prime\prime}})[\delta(E_{A_{2}}^{0}+\mathcal{E}_{A_{2}}%
^{b}-E_{B_{1}}+\hbar\omega_{\alpha^{\prime\prime}})+\delta(E_{A_{2}}%
^{0}+\mathcal{E}_{A_{2}}^{b}-E_{B_{1}}-\hbar\omega_{\alpha^{\prime\prime}})]
\]%
\[
+\frac{1}{4}\sum_{\alpha^{\prime}\alpha^{\prime\prime}}[\frac{(K_{A_{2}B_{1}%
}^{\prime\alpha})^{2}\hbar}{2M_{\alpha^{\prime}}\omega_{\alpha^{\prime}}%
}+\frac{(K_{A_{2}B_{1}}^{\prime\alpha}\Theta_{\alpha^{\prime}}^{A_{2}})^{2}%
}{4}\coth\frac{\beta\hbar\omega_{\alpha^{\prime}}}{2}]\text{csch}\frac
{\beta\hbar\omega_{\alpha^{\prime}}}{2}f_{EL}(\omega_{\alpha^{\prime\prime}})
\]%
\begin{align*}
\times &  [\delta(E_{A_{2}}^{0}+\mathcal{E}_{A_{2}}^{b}-E_{B_{1}}+\hbar
\omega_{\alpha^{\prime}}+\hbar\omega_{\alpha^{\prime\prime}})+\delta(E_{A_{2}%
}^{0}+\mathcal{E}_{A_{2}}^{b}-E_{B_{1}}+\hbar\omega_{\alpha^{\prime}}%
-\hbar\omega_{\alpha^{\prime\prime}})\\
+  &  \delta(E_{A_{2}}^{0}+\mathcal{E}_{A_{2}}^{b}-E_{B_{1}}-\hbar
\omega_{\alpha^{\prime}}+\hbar\omega_{\alpha^{\prime\prime}})+\delta(E_{A_{2}%
}^{0}+\mathcal{E}_{A_{2}}^{b}-E_{B_{1}}-\hbar\omega_{\alpha^{\prime}}%
-\hbar\omega_{\alpha^{\prime\prime}})]
\end{align*}%
\[
+[\sum_{\alpha^{\prime}}\frac{(K_{A_{2}B_{1}}^{\prime\alpha}\Theta
_{\alpha^{\prime}}^{A_{2}})^{2}}{64}(2-\text{csch}^{2}\frac{\beta\hbar
\omega_{\alpha^{\prime}}}{2})]\sum_{\alpha^{\prime\prime}\alpha^{\prime
\prime\prime}}f_{EL}(\omega_{\alpha^{\prime\prime}})f_{EL}(\omega
_{\alpha^{\prime\prime\prime}})
\]%
\[
\times\lbrack\delta(E_{A_{2}}^{0}+\mathcal{E}_{A_{2}}^{b}-E_{B_{1}}%
+\hbar\omega_{\alpha^{\prime\prime}}+\hbar\omega_{\alpha^{\prime\prime\prime}%
})+\delta(E_{A_{2}}^{0}+\mathcal{E}_{A_{2}}^{b}-E_{B_{1}}+\hbar\omega
_{\alpha^{\prime\prime}}-\hbar\omega_{\alpha^{\prime\prime\prime}})
\]%
\[
+\delta(E_{A_{2}}^{0}+\mathcal{E}_{A_{2}}^{b}-E_{B_{1}}-\hbar\omega
_{\alpha^{\prime\prime}}+\hbar\omega_{\alpha^{\prime\prime\prime}}%
)+\delta(E_{A_{2}}^{0}+\mathcal{E}_{A_{2}}^{b}-E_{B_{1}}-\hbar\omega
_{\alpha^{\prime\prime}}-\hbar\omega_{\alpha^{\prime\prime\prime}})]
\]%
\[
-\sum_{\alpha^{\prime}}\frac{(K_{A_{2}B_{1}}^{\prime\alpha}\Theta
_{\alpha^{\prime}}^{A_{2}})^{2}}{16}\text{csch}^{2}\frac{\beta\hbar
\omega_{\alpha^{\prime}}}{2}[\delta(E_{A_{2}}^{0}+\mathcal{E}_{A_{2}}%
^{b}-E_{B_{1}}+2\hbar\omega_{\alpha^{\prime}})+\delta(E_{A_{2}}^{0}%
+\mathcal{E}_{A_{2}}^{b}-E_{B_{1}}-2\hbar\omega_{\alpha^{\prime}})]+\cdots\}
\]
where%
\begin{equation}
f_{EL}(\omega_{\alpha})=\frac{1}{2}(\theta_{\alpha}^{A_{2}})^{2}%
\text{csch}\frac{\beta\hbar\omega_{\alpha}}{2} \label{fel}%
\end{equation}
In conduction band, the energy of any localized state is lower than that of
any extended state. Thus zero-phonon process cannot conserve energy and
therefore\ is not possible. Because $\Delta G_{EL}^{0}<0,$ factor
$\exp[|\Delta G_{EL}^{0}|/(2k_{B}T)]$ increases with decreasing temperature.
On the other hand, other factors in Eq.(\ref{loweL}) decrease with lowering
temperature. There exists an optimal temperature T$_{\ast}$, at which
transition probability is maximum. In non-equilibrium phenomenon like
luminescence, \ T$_{\ast}$ may have some traces.

\section{transition between extended states caused by electron-phonon
interaction}

If in the initial moment the \textquotedblleft one electron+ many
nuclei\textquotedblright\ system is at state $|B_{1}\cdots N_{\alpha}^{\prime
}\cdots\rangle$, then only $F^{B_{1}}(\cdots N_{\alpha}^{\prime}\cdots;t=0)=1$
and other coefficients are zero. At later moment $t$, probability amplitude
$F^{B_{2}}(\cdots N_{\alpha}\cdots;t)$ is determined by the first order
perturbation theory based on Eq.(\ref{fq}):%
\begin{equation}
i\hbar\frac{\partial F_{\{N_{\alpha}\}}^{B_{2}}}{\partial t}=\langle
B_{2}\cdots N_{\alpha}\cdots|V_{EE}^{e-ph}|B_{1}\cdots N_{\alpha}^{\prime
}\cdots\rangle F_{\{N_{\alpha}^{\prime}\}}^{B_{1}}e^{it(\mathcal{E}_{B_{2}%
}^{\{N_{\alpha}\}}-\mathcal{E}_{B_{1}}^{\{N_{\alpha}^{\prime}\}})/\hbar}
\label{EEP}%
\end{equation}
The matrix element (\ref{ee1}) an be calculated from the recursion relation
for $z\Phi_{N}(z)$. The transition probability from extended state
$|B_{1}\rangle$ to extended state $|B_{2}\rangle$ is%

\begin{equation}
W(B_{1}\rightarrow B_{2})=\frac{2\pi}{\hbar}\sum_{\alpha^{\prime}}%
(K_{B_{2}B_{1}}^{\alpha^{\prime}})^{2}\frac{\hbar}{M_{\alpha^{\prime}}%
\omega_{\alpha^{\prime}}}[\frac{\overline{N_{\alpha^{\prime}}^{\prime}}}%
{2}\delta(E_{B_{2}}-E_{B_{1}}-\hbar\omega_{\alpha^{\prime}})+\frac
{\overline{N_{\alpha^{\prime}}^{\prime}}+1}{2}\delta(E_{B_{2}}-E_{B_{1}}%
+\hbar\omega_{\alpha^{\prime}})]\label{avie}%
\end{equation}
where $\overline{N_{\alpha^{\prime}}^{\prime}}=(e^{\beta\hbar\omega
_{\alpha^{\prime}}}-1)^{-1}$ is the average phonon number in the
$\alpha^{\prime}$th mode. In a crystal, Eq.(\ref{avie}) arises from inelastic
scattering with phonons. Although in amorphous solids one cannot classify the
electronic states and the phonon states with wave vectors, Eq.(\ref{avie}) is
similar to its crystalline counterpart of phonon inelastic scattering. The
relaxation time $\tau(B_{1})$ of state $|B_{1}\rangle$ can be defined as
\begin{equation}
\frac{1}{\tau(B_{1})}=\sum_{B_{2}}W(B_{1}\rightarrow B_{2})[1-f(B_{2}%
)]\label{rti}%
\end{equation}%
\begin{equation}
\sigma^{EE}=n_{E}e\mu^{EE},\text{ \ \ \ }\mu^{EE}=\frac{e\tau}{m}\label{mdri}%
\end{equation}
where $n_{E}$ is the number of electrons in extended states, $\mu^{EE}$ is the
mobility of the electrons in extended states. From the observed
value\cite{cer} of $\mu_{\text{drift}}=0.61$cm$^{2}$V$^{-1}$Sec$^{-1}$ of
a-Si, one deduces relaxation time $\tau\thicksim0.35$fs. Take effective
nuclear charge Z$^{\ast}$ of Si as 4, bond length $b=$2.35\AA , the coupling
constant will be $K\thicksim\frac{Z^{\ast}e^{2}}{4\pi\epsilon_{0}(b/2)^{2}%
}\thicksim9.2\times10^{-8}$N.  $W\thicksim\frac{K^{2}}{\hbar M\omega^{2}%
}\thicksim6.7\times10^{12}$sec$^{-1}$. The relaxation time $\tau
=W^{-1}\thicksim0.15$ps. The electron moving in extended states scattered by
phonons will give a mobility 3 order of magnitude larger than the observed
value of `drift mobility'. Therefore the drift of electron in extended states
is not responsible for the observed `drift mobility'.

\subsection{Four types of transitions and conduction mechanisms}

Some features of 4 types of transitions are summarized in TABLE
\ref{tab:table1}. With the basic transition probabilities for four types
transitions in Sec.III-Sec.VI, the transport properties could be derived from
various master equations (Fokker-Planck equation, Boltzmann equation etc.), we
will not further pursue this well-known but complicated procedure and only
briefly discuss\ the conduction mechanisms.

In thermal equilibrium at temperature $T$, the ratio $R$ of the number of
electrons in the extended states to the number of electrons in localized
states is of order $e^{-F/k_{B}T}$, $F$ is the distance between the lower
mobility edge and the bottom of conduction band (for a-Si, $F$ is
about\cite{ster} $0.2$eV. At T=300K, $R\thicksim4\times10^{-4}$). However the
mobility for a carrier in extended states is much higher than the mobility in
localized states. For the extended states outside the short lifetime belt, the
life time $t_{\text{life}}$\ of a carrier in one of these states is\ long
enough. Under the influence of external electric field, the carrier drifts in
these extended states. Thus at higher temperature, transition from localized
states to extended states (LE) will lead to an appreciable contribution to
conductivity $\sigma_{dc}^{(2)}=\frac{e^{2}n_{E}}{m}\frac{\tau t_{\text{life}%
}}{\tau+t_{\text{life}}}$,where $m$ is mass of electron, $n_{E}$ is the number
of electrons in extended \ states. $\tau\thicksim\lbrack W(B_{1}\rightarrow
B_{2})]^{-1}$ is the mean free time in the extended states determined by
Eq.(\ref{avie}). When $t_{\text{life}}>>\tau$, $\sigma_{dc}^{(2)}$ is reduced
to the usual formula $\tau e^{2}n_{E}/m$.

In present perturbation treatment, the total conduction comes from four
processes: (1) the hopping among localized states; (2) the hoping from a
localized state to an extended state; (3) the hopping from an extended state
to a localized state and (4) the drift in extended state scattered by phonon.
The carriers are pumped to the extended states by type (2) transition.

\begin{table}[ptb]
\caption{Some features of 4 types of transitions}%
\label{tab:table1}%
\begin{ruledtabular}
\begin{tabular}{ccccccccccccc}
transitions & origin & phonons needed&  activated & role in
conduction & $W_{T}$(sec$^{-1}$) & \\
\hline
L$\rightarrow$L & J & multi&  \ \ \ yes &
direct &  $10^{12}$& \\
\hline
L$\rightarrow$E & $J^{'}$ & multi&  \ \ \ yes & direct+indirect
& $10^{10}-10^{13}$ & \\
\hline
E$\rightarrow$L & $K^{'}$ & multi&  \ \ \ yes &
direct+indirect &$10^{13}-10^{14}$  & \\
\hline
E$\rightarrow$E & $K$ & single&  \ \ \ no & reduce &$10^{13}$
& \\
\end{tabular}
\end{ruledtabular}
\end{table}

The last column gives the order of magnitude of the transition probability
estimated for a-Si at T=300K. The value of LL transition is for the transition
between two nearest neighbors. For an intrinsic or a lightly n-doped
semiconductor at not too high temperature (T%
$<$%
580K for a-Si), only the lower part of conduction tail is occupied. $\Delta
G_{LE}$ is large, LE transition probability is about two order of magnitude
smaller than that of LL\ transition. For an intrinsic semiconductor at higher
temperature or a doped material, $\Delta G_{LE}^{0}$ becomes comparable with
$\Delta G_{LL}^{0}$, LE transition probability is about ten times larger than
that of LL transition. The first three types of transitions increase the
mobility of an electron, whereas type (4) transition decreases mobility of an
electron. Although a stricter solution of Eq.(\ref{fs1}) could be obtained by
considering the change in the functional form of vibrational states by the
motion of the electrons and the change in electronic wave function in the same
time. However for a localized state in amorphous semiconductor, the time spent
from one localized state to another (10$^{-10}$m/10$^{5}$m$\cdot$%
sec$^{-1}\thicksim10^{-16}-10^{-15}$sec) is much shorter than the stay time on
one localized state ($10^{-12}$sec). In addition the slow multi-phonon
reorganization process of a configuration reduces the impulse of\ an electron
on the network when the electron moves. Thus one can neglect the recoil of
lattice by moving an electron. This differs from an ionic crystal where the
deformation of the lattice closely stalks the motion of the slow electron.

\subsection{Long time and higher order processes}

The perturbation treatments of the fundamental processes given in
Sec.IV-Sec.VII are only suitable for short times, in which the probability
amplitude of the final state is small. Starting from a localized state we only
have $L\rightarrow E$ process and $L\rightarrow L$ process. Starting from an
extended state, we\ only have $E\rightarrow L$ process and $E\rightarrow E$
process. For long time period, higher order processes appear. For example
$L\rightarrow E\rightarrow L\rightarrow L\rightarrow E\rightarrow L\rightarrow
E\rightarrow E\rightarrow L$ etc. Those high order processes are what occur in
the amorphous solids. However, the perturbation picture could be used to
compute transport properties. In semiconductors, the number of carriers are
smaller than the number of avialable\ states. The motions of the individual
carriers can be viewed as independent. On the other hand, for macroscopic
sample, each type of states are highly populated if we consider the whole
sample. Thus if we concern the collective behavior of all carriers rather than
one individual carrier in a long time period, the picture of the four types of
transitions works well statistically.

\section{Summary}

For amorphous solids, following Holstein's work on small polaron, we
established the evolution equations for localized states and extended states
in the presence of lattice vibrations (Eqs.(\ref{s1}) and (\ref{s2})). To
simplify the evolution equations, for localized states, we restrict ourselves
to the most localized ones. These localized states are attached to the most
distorted regions, and are spatially well separated. One can neglect the
overlap integral\ between two of them and the overlap integral between an
extended state and one of the most localized states. For any process involving
localized states close to mobility edge, the conclusions obtained from the
simplified evolution equations (Eqs.(\ref{s11}) and (\ref{s22})) are only
qualitatively correct.

For short times ($t<<\frac{\hbar}{J},$ $\frac{\hbar}{J^{\prime}},$
$\frac{\hbar}{K},$ $\frac{\hbar}{K^{\prime}}$), perturbation theory can be
used to solve Eqs.(\ref{s11}) and (\ref{s22}). One obtains the transition
probabilities of LL, LE, EL and EE transitions. In high temperature
($k_{B}T\gtrsim\hbar\overline{\omega}$) and low temperature ($k_{B}%
T<\hbar\overline{\omega}/10$) limits, the `time' integral in transition
probabilities of LL, LE and EL transitions can be carried out analytically.
The relative errors are less than 10$^{-3}$. In amorphous semiconductors (e.g.
a-Si), the transfer integrals $J$ and $J^{\prime},$ and the e-ph interactions
$K^{\prime}$ and $K$ are sufficiently small such that the perturbation theory
is applicable for a meaningful time period. Although in a sample the motion of
a single carrier is a long time and high order process. Because conduction is
a collective behavior of many carriers, the results of perturbation theory can
be used.

The hopping motion of electron along the direction of electric field during
LL, LE and EL transitions directly contributes to conductivity. In external
electric field, an electron in extended states moves along the direction of
electric field. EE transition deviates the direction of drift which is along
the direction of field and reduces the transport of charges. LE transition
increases the number of electrons in extended states, whereas EL transition
decreases the number of electrons in extended states. They also affect
conductivity indirectly.

At `very' high temperature ( $k_{B}T\gtrsim2.5\hbar\overline{\omega}$), the
transition probabilities for the transitions of type (1), type (2) and type
(3) are reduced to an Eq.(\ref{ea})-like formula. It is universal for the
processes involving localized state. \ If a localized state close to the
bottom of conduction band, $\Delta G_{LE}-\Delta G_{LL}$ is about the mobility
edge, $W^{LE}$ is one or two order of magnitude smaller than $W^{LL}$. If a
localized state close to mobility edge, $\Delta G_{LE}\thicksim\Delta G_{LL},$
because $J^{\prime}$ is several times larger than $J,$ $W^{LE}$ could be one
order of magnitude larger than $W^{LL}$. The probability of type (3) is one or
two order of magnitude larger than that of LL transition. The reason is
$\Delta G_{EL}<0$, $E_{a}^{EL}$ is smaller than $E_{a}^{LL}$ while $K^{\prime
}\Theta$ is same order of magnitude as $J$. The probability of EE transition
is about 10$^{3}$ times larger than $W^{LL}$ (cf. Table I). Type (4)
transition deviates the direction of the drift motion along the direction of
electric field and reduces conductivity. This is contrast with LL, LE and EL
transition. The relative contribution to conductivity of four types of
transition also highly depends on the number of\ carriers in extended states
and the number of carriers in localized states, which are determined by the
extent of doping and temperature.

At low temperature ( $k_{B}T<\hbar\overline{\omega}/10$), the non-diagonal
transition is still multi-phonon activated process whatever it is LL, LE or EL
transition. The activation energy is just half of the energy difference
between final state and initial state ($\Delta G_{LL}/2,$ $\Delta G_{LE}/2,$
$\Delta G_{EL}/2$). The reorganization of vibration configuration is not
needed, this is similar to the hopping between mid-gap states\cite{motv,aps}
which does not need reorganization of the static displacements\cite{md}. Thus
we expect variable range hopping also works for the LL transitions which are
transitions between tail states.

In a-Si for both high and low temperature regimes, the computed temperature
dependence of mobility of LL transition agrees with the
observed\cite{mo,tie,cer,koc} `hopping mobility'. EE transition offers a
mobility which is three times larger than the observed drift mobility.
However, in an intrinsic or a lightly doped system, the occupation probability
of the extended states is small. The observed `drift mobility' may come from
LE transition.

There exists a short lifetime belt of the extended states inside conduction
band or valence band. These states favor non-radiation transition by emitting
several phonons, have smaller contributions to dc conductivity than other
extended states. The fact that photoluminescence lifetime in a-Si/SiO$_{2}$
decreases with frequency confirms the existence of short lifetime belt in
conduction band. The phonon-assisted non-radiation transition dissipates the
population of the excited states in amorphous semiconductors.

Some important points are not addressed in this work. The overlap integral
between two less localized states which spread in several distorted regions
and the overlap integral between one less localized state and an extended
state require the wave function of a less localized state which cannot be
described by a single exponential decay function. The coherence between states
is subtle in Eqs.(\ref{1s}) and (\ref{2s}). The dependence of conduction on
the strength of external field also requires including electric field in the
zeroth order in the evolution equations. The drag effect\cite{con06} of water
on the conduction of solvated DNA is not included in Eq.(\ref{fs1}). We wish
to discuss them in near future.

\section{Acknowledgements}

We thank the Army Research Office for support under MURI W91NF-06-2-0026, and
the National Science Foundation for support under grants DMR 0600073 and
0605890. DAD thanks the Leverhulme Trust (UK) and the National Science
Foundation for sabbatical support.

\end{document}